\documentclass[aps,prl,reprint]{revtex4-2}
\usepackage{blindtext}
\usepackage{graphicx}
\usepackage{amsmath}
\usepackage{upgreek}
\usepackage[english]{babel}

\makeatletter
\let\ORIbbl@fixname\bbl@fixname
\def\bbl@fixname#1{%
  \@ifundefined{languagealias@\expandafter\string#1}
    {\ORIbbl@fixname#1}
    {\edef\languagename{\@nameuse{languagealias@#1}}}%
}
\newcommand{\definelanguagealias}[2]{%
  \@namedef{languagealias@#1}{#2}%
}
\makeatother

\definelanguagealias{en}{english}
\definelanguagealias{eng}{english}

\usepackage{braket}
\usepackage{hyperref}
\hypersetup{colorlinks=true,linkcolor=blue, filecolor=blue, urlcolor=blue, citecolor=blue}

\begin{document}
\title{Tunable Gyromagnetic Augmentation of Nuclear Spins in Diamond}
\author{R. M. Goldblatt}
\author{A. M. Martin}
\author{A. A. Wood}
\email{alexander.wood@unimelb.edu.au}
\affiliation{School of Physics, University of Melbourne, Parkville Victoria 3010, Australia}
\date{\today}

\begin{abstract}
Nuclear spins in solids exhibit long coherence times due to the small nuclear gyromagnetic ratio. This weak environmental coupling comes at the expense of slow quantum gate operations, which should be as fast as possible for many applications in quantum information processing and sensing. In this work, we use nitrogen-vacancy (NV) centers in diamond to probe the nuclear spins within dark paramagnetic nitrogen defects (P1 centers) in the diamond lattice. The gyromagnetic ratio of the P1 nuclear spin is augmented by hyperfine coupling to the electron spin, resulting in greatly enhanced coupling to radiofrequency control fields. We then demonstrate that this effect can be tuned by variation of an external magnetic field. Our work identifies regimes in which we are able to implement fast quantum control of dark nuclear spins, and lays the foundations for further inquiry into rapid control of long-lived spin qubits at room temperature. 
\end{abstract}

\maketitle
Nuclear spins in solid-state substrates have attracted considerable interest for applications in quantum information processing \cite{fuchs_quantum_2011, pla_coherent_2014, hensen_silicon_2020, muhonen_storing_2014, degen_quantum_2017} and quantum sensing \cite{maclaurin_measurable_2012, ledbetter_gyroscopes_2012, ajoy_stable_2012, liu_nanoscale_2019, soshenko_nuclear_2021, jarmola_demonstration_2021}, as they exhibit long coherence times at room temperature \cite{kane_silicon-based_1998, dutt_quantum_2007, ladd_quantum_2010, maurer_room-temperature_2012, zhong_optically_2015}. However, reduced coupling to magnetic noise comes with concomitantly weaker coupling to control fields, meaning gate operations are slow and error-prone.   

When coupled to optically active electron spins, such as the NV center in diamond \cite{doherty_nitrogen-vacancy_2013, schirhagl_nitrogen-vacancy_2014, wu_diamond_2016}, nuclear spins can be dynamically polarized~\cite{jacques_dynamic_2009}, controlled and measured, even down to the single-spin level ~\cite{neumann_single-shot_2010}, via hyperfine coupling to the electron spin. Nuclear spins coupled to an electron spin via a spin-mixing hyperfine interaction exhibit an enhanced coupling to resonant driving fields \cite{smeltzer_robust_2009, sangtawesin_hyperfine-enhanced_2016, degen_entanglement_2021, chen_measurement_2015, sangtawesin_quantum_2016, goldman_optical_2020}. This gyromagnetic augmentation has been demonstrated for nearby $^{13}$C nuclei ~\cite{childress_coherent_2006} and the intrinsic $^{14}$N nuclear spin hosted within the NV defect, enabling fast manipulation of nuclear spins \cite{chen_measurement_2015, sangtawesin_hyperfine-enhanced_2016} while still preserving the longer coherence times relative to the electron spin \cite{jarmola_robust_2020, wood_quantum_2021}. Gyromagnetic augmentation offers a means of tuning a qubit between regimes of strong and weak interaction, while still retaining long coherence times.

More recently, substitutional nitrogen atoms in diamond (P1 centers), composed of a nitrogen nucleus coupled to an electron spin, have been identified as a promising platform for quantum information processing \cite{laraoui_nitrogen-vacancy-assisted_2012, belthangady_dressed-state_2013, knowles_demonstration_2016}.  The ability to control and entangle individual electron and nuclear spins of P1 centers has been demonstrated at cryogenic temperatures \cite{degen_entanglement_2021}. In NV quantum sensing, magnetic noise from the P1 spin bath limits the coherence of the NV itself, particularly in diamonds with high nitrogen density ~\cite{de_lange_controlling_2012, knowles_observing_2014, bauch_decoherence_2020}. Driving the P1 bath using resonant radiofrequency (rf) fields~\cite{de_lange_universal_2010, de_lange_controlling_2012, knowles_observing_2014, hansom_environment-assisted_2014, bauch_ultralong_2018} has been demonstrated to mitigate P1 dephasing and preserve NV coherence. Such continuous dynamical decoupling also requires detailed knowledge of the spin bath frequency spectrum, which changes significantly with magnetic field.

In this work, we use an ensemble of NV centers in diamond to experimentally characterize the electronic augmentation of optically-dark P1 nuclear spins in diamond, and demonstrate rapid control of the nuclear spin state at room-temperature. We show that the augmentation of the nuclear gyromagnetic ratio is dependent on the external magnetic field strength, enabling fast quantum gate operations on the P1 nuclear spins at low magnetic field strength where spin mixing is significant, and explaining the vanishing of P1 nuclear spin effects at higher field. Detailed characterization of the P1 bath has implications for schemes using P1 qubits in quantum information processing and harnessing continuous dynamical decoupling to eliminate P1-induced NV dephasing. 

A schematic of our experiment is depicted in Figure \ref{exp1} and is described in greater detail in refs. \cite{wood_anisotropic_2021} and \footnote{See Supplementary Information.}. A type 1b, $\langle 111 \rangle$-cut diamond sample with a $1.1\%$ natural abundance of $^{13}$C and 1\,ppm N concentration is mounted on the spindle of an electric motor, which acts in this case as a precision rotation stage. Current-carrying coils are used to generate a variable magnetic field of up to $100$ G aligned along the $\hat{z}$-axis. Microwaves for NV driving and rf for P1 control are produced by wires with diameters $20\,\upmu$m and $50\upmu$m respectively, arranged in a cross as shown in Fig. \ref{exp1}(a). The required rf pulses are produced by an I/Q modulated vector signal generator \footnote{See Supplementary information}. We use sample where coherence times are dominated by $^{13}$C spins, instead of a nitrogen dominated sample, so that the coherence time of the NV ensemble is long enough to allow observation of slow (several microsecond) Rabi oscillations.

The P1 center features an electron spin ($S$ = 1/2) associated with an unpaired electron, coupled to an $I$ = 1 $^{14}$N nucleus (99.6\% nat. abundance). The P1 center exhibits a static Jahn-Teller (JT) distortion \cite{davies_dynamic_1979, davies_jahn-teller_1981,ammerlaan_reorientation_1981}, which results in four possible crystallographic orientations for the P1 quantization axis due to an elongation of one of the N-C bonds, as shown in Fig. \ref{exp1}(c). The interaction Hamiltonian for a single P1 center is given by:
\begin{align}
\begin{split} \label{HP1}
H_{P1} ={}& -\gamma_e \vec{\boldsymbol{B}} \cdot \vec{\boldsymbol{S}} \, - \, \gamma_N \vec{\boldsymbol{B}} \cdot \vec{\boldsymbol{I}} + \, A_{\parallel} S_z I_z \\
& + A_{\perp} (S_x I_x + S_y I_y) \, + \,  Q I_z^2 \, ,
\end{split}
\end{align}
where $\gamma_e / 2\pi = -2.8$ MHz/G and $\gamma_N / 2\pi = 307.7$ Hz/G are the gyromagnetic ratios of the electron and of the $^{14}$N nuclear spin, respectively. In Eq. \ref{HP1}, the hyperfine interaction between the electron spin and the $^{14}$N nucleus has been separated into an axial coupling term, $A_{\parallel} / 2\pi = 114$ MHz, and a transverse component, $A_{\perp} / 2\pi = 81.34$ MHz. The nuclear quadrupole coupling term, with $Q / 2\pi = -4.2$ MHz, defines the zero-field splitting between the $m_I = \pm 1$ and $m_I = 0$ states. Also, $\vec{\boldsymbol{S}} = (S_x, S_y, S_z)$ and $\vec{\boldsymbol{I}} = (I_x, I_y, I_z)$ are the electron and nuclear spin operators of the P1 center, and $\vec{\boldsymbol{B}}$ is an applied magnetic field. 

\begin{figure}[tbp]
\center	\includegraphics[width=0.5\textwidth,keepaspectratio, trim = {1cm 1cm 0 1cm}]{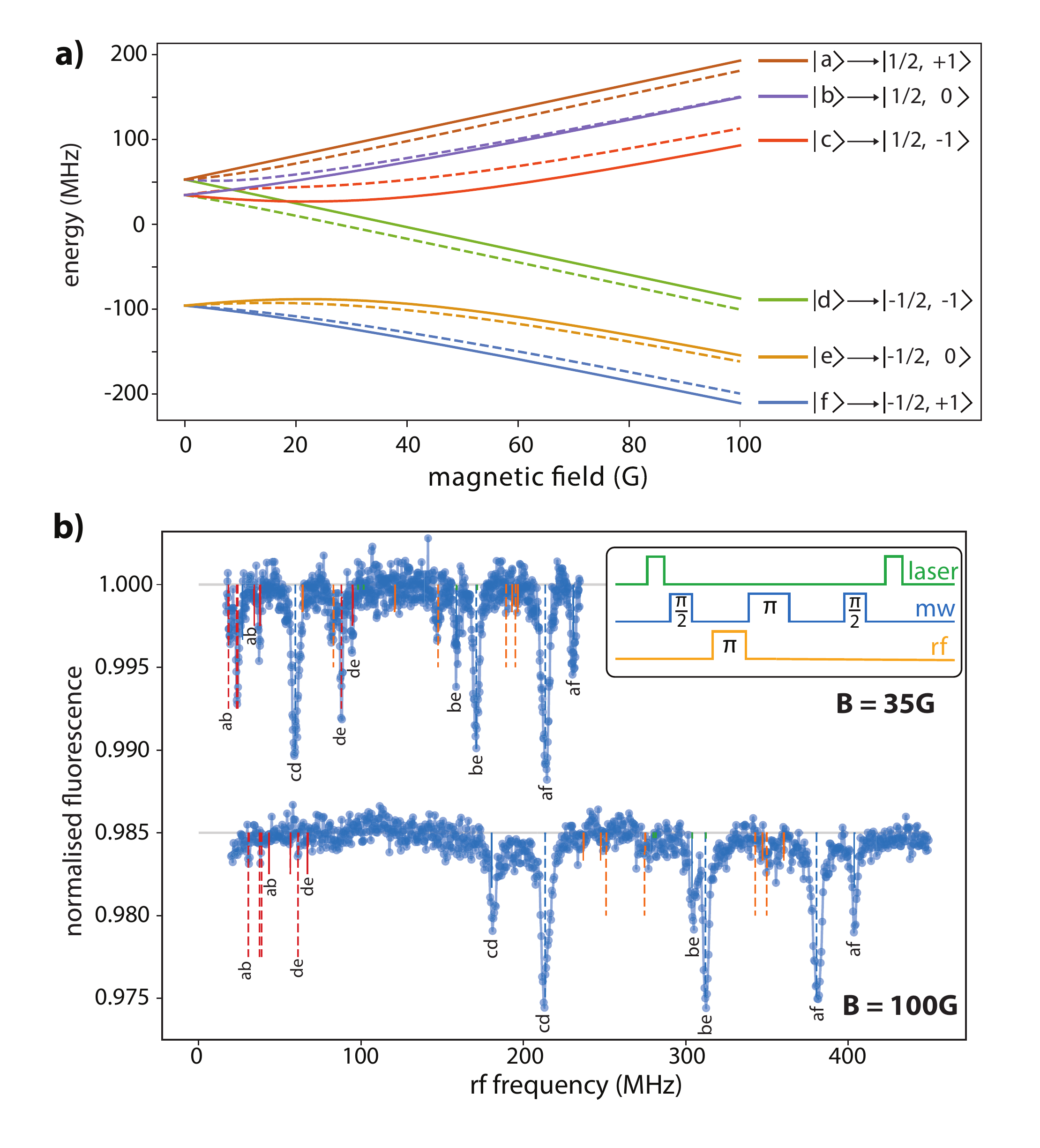}
	\caption{\label{deers2} Energy levels and DEER measurements for the P1 spin bath. (a) Calculated energy levels for the coupled P1 electron spin and $^{14}$N nuclear spin as a function of magnetic field. The solid (dashed) lines represent the energy levels for the on-axis (off-axis) P1 orientation classes. The six energy levels for each orientation class are labeled from $\ket{a}$ to $\ket{f}$, according to their electron and nuclear spin state in the limit $B \rightarrow \infty$. (b) DEER measurements and  pulse sequence (inset). A spin-echo sequence is applied to the NV centers, while an rf $\pi$-pulse with swept frequency is applied at a constant time in the sequence. DEER measurements at 35\,G (top) and 100\,G (bottom). The calculated resonant frequencies for particular transitions are displayed over the experimental signal, color-coded by transition type: blue for single electron spin transitions, yellow for double quantum transitions and red for nuclear spin transitions.}
	\vspace{-0.3cm}
\end{figure}

We assume the magnetic field is aligned approximately along a particular NV crystallographic orientation class (and, therefore, one of the P1 orientation axes), which we define as the $\hat{z}$-axis. The magnetic field has therefore a polar angle $\theta$ to the $\hat{z}$-axis, such that $\theta = 0^{\circ}$ for the P1 orientation class parallel to the magnetic field, which will be referred to as `on-axis' and $\theta = 109.5^{\circ}$, for the other three degenerate P1 orientations,  denoted `off-axis'. The energies for the six P1 spin states are calculated from the Hamiltonian in Eq. \ref{HP1} and plotted as a function of magnetic field strength in Fig. \ref{deers2}(a) for both the on-axis and off-axis orientation classes. In the high magnetic field limit, $m_S = -1/2, +1/2$ and $m_I = -1, 0, +1$ are good quantum numbers, and we label the eigenstates of $H_{P1}$ as $\ket{m_S,m_I}$. At lower magnetic fields, the eigenstates of the Hamiltonian are superpositions of electronic and nuclear spin states. To account for this state-mixing at low fields, we identify the asymptotic eigenstates in the high field limit and label the states $\ket{a} - \ket{f}$ in descending order as shown in Fig. \ref{deers2}(a). 

\begin{figure}[tbp]	\center{\includegraphics[width=0.5\textwidth,keepaspectratio, trim = {0.5cm 1cm 0 0.5cm}]{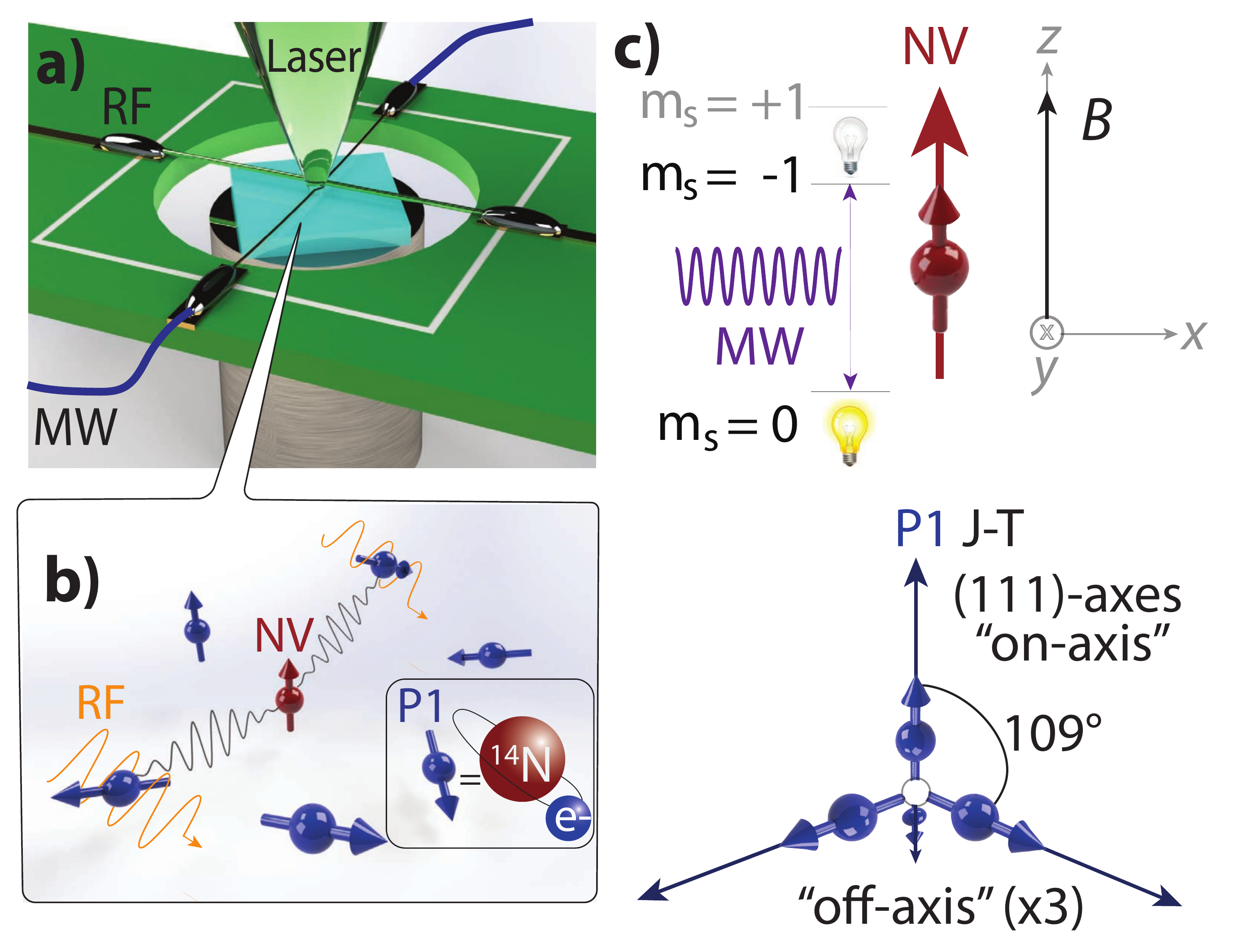}} 
	\caption{\label{exp1} Experimental configuration and system geometry. a) Microwave and radiofrequency fields are generated by crossed wires to control the NV and P1 centers. The ensemble of NVs are optically prepared and read out by a confocal microscope using a 532nm laser. b) The NV centers interact with surrounding P1s, which exhibit a Jahn-Teller distortion that gives rise to four orientation classes (inset). c) The energy levels of the NV center and the orientations of the P1 centers with respect to the external magnetic field, which is aligned to the $\langle 111 \rangle$ NV axis.}
	\vspace{-0.5cm}
\end{figure}

We use double electron-electron resonance (DEER) spectroscopy \cite{larsen_double_1993} to characterize the dark spin bath surrounding the NV centers, as depicted in Fig. \ref{deers2} (b). A spin-echo pulse sequence decouples the NV ensemble from its quasistatic environment, with the free evolution time of the sequence fixed at a $^{13}$C-induced revival time in order to ensure signal visibility~\cite{childress_coherent_2006}. An rf $\pi$-pulse recouples resonant spins in the environment to the NV centers as we sweep the frequency of the rf field. The results from a DEER experiment at $B =$100\,G are shown at the bottom of Fig. \ref{deers2} (b). The expected frequencies for all possible P1 spin transitions, calculated from Eq. (\ref{HP1}), are also shown. We observe six spectral features, which are consistent with the electron spin transitions in the theoretical spectrum (i.e. $\Delta m_S = \pm 1$ and $\Delta m_I = 0$). For both theory and experiment, the deeper amplitudes of the resonant peaks correspond to the off-axis P1s, due to increased contrast resulting from the three degenerate orientations. 

The DEER measurement was then repeated at a magnetic field of 35\,G, as shown at the top of Fig. \ref{deers2} (b), in which we see the same electron spin transitions. We also see additional spectral features, with comparable amplitudes, corresponding to nuclear spin transitions in which the nuclear spin projection changes ($\Delta m_I = \pm 1$) while the electron spin is conserved ($\Delta m_S = 0$). At a magnetic field of 100\,G, the nuclear spin transition features are indistinguishable from the background of the spectrum. Nuclear spectral features were also absent from previous DEER measurements at comparable magnetic fields \cite{de_lange_controlling_2012, knowles_demonstration_2016, bauch_ultralong_2018, pagliero_multispin-assisted_2018}, which suggests that the coupling of the P1 nuclear spin to the rf field depends on the magnitude of the external magnetic field. 

To better understand the variation in rf coupling, we consider how the nuclear spin is coupled to the P1 electron spin. The augmentation factor, $\alpha \equiv \gamma_{\text{N,eff}} / \gamma_{\text{N,bare}}$, has been previously measured for a $^{14}$N nuclear spin coupled to the NV electron spin in the presence of a magnetic field that is parallel to the N-V axis \cite{chen_measurement_2015, sangtawesin_quantum_2016} and recently for off-axis field in the same system~\cite{wood_quantum_2021}. Here, the derivation of the augmentation factor is extended to the coupled electron and nuclear spin of the P1 center, and includes the case in which the magnetic field is not aligned with the electron spin quantization axis in order to account for the different orientation classes of the P1 ensemble. 

The contribution of the transverse hyperfine coupling, parametrized by $A_{\perp}$, dominates the P1 Hamiltonian at low magnetic fields. Hence, the effective nuclear gyromagnetic ratio is expected to increase towards lower field, where state mixing is most apparent. The amplitude of the DEER signal is indicative of the strength of the dipolar coupling between the NV and the P1 spin. As the effective nuclear gyromagnetic ratio reduces with increasing field, so does the amplitude of the DEER signal, as we observe in the difference between DEER measurements at 35\,G and 100\,G, presented in Fig. \ref{deers2}. Following Ref. \cite{chen_measurement_2015}, the Hamiltonian in Eq. \ref{HP1} can be diagonalized using $E = P^{-1} \hat{H} P$, where $E$ is a diagonal matrix of energy eigenvalues and $P$ is the matrix of the eigenvectors of the Hamiltonian. We compute the matrix elements describing the coupling between states in the  basis where $\hat{H}$ is diagonal by transforming the rf Hamiltonian, $\hat{H}_\text{rf} = B_\text{rf}(t) (\gamma_E S_{x} + \gamma_{N}  I_{x})$, into the diagonal basis. The time-dependence of $\hat{H}_\text{rf}$ is eliminated using the rotating-wave approximation. We can then compute the augmentation factor for a given nuclear spin transition, that is:
\begin{equation} \label{alpha}
\alpha_{a,b} = \frac{1}{\gamma_N B_\text{rf}} \bra{a} \tilde{H}_\text{rf} \ket{b}
\end{equation}
for transition $\ket{a} \rightarrow \ket{b}$, where $B_\text{rf}$ is the amplitude of the oscillating field. The $\tilde{H}_\text{rf}$ term denotes the interaction Hamiltonian in the diagonal basis defined by $P$, which implicitly contains within it the field-dependence of this expression. 


\begin{figure}[tbp]	\center{\includegraphics[width=0.5\textwidth, keepaspectratio, trim = {0.5cm 0.5cm 0 1cm}]{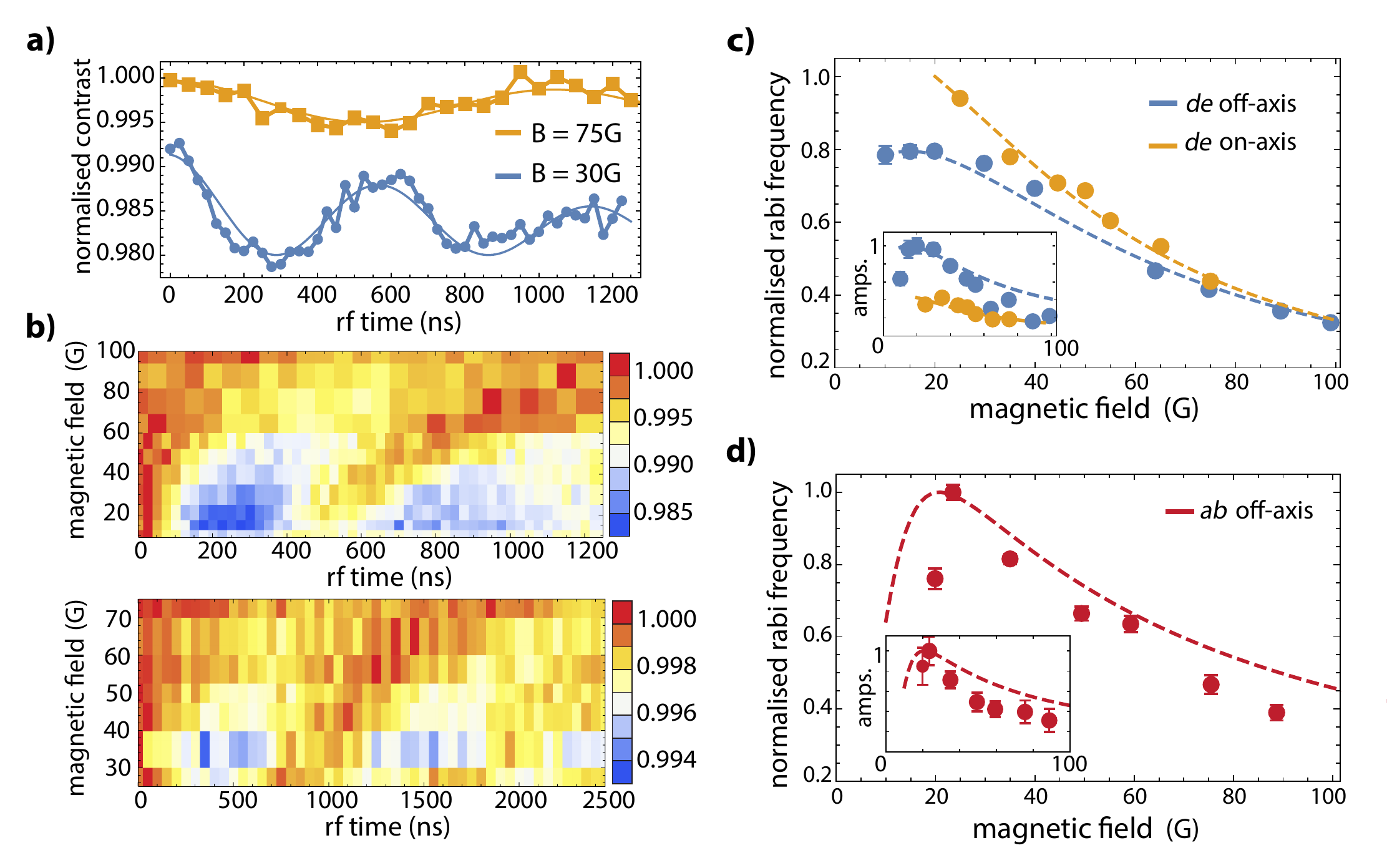}} 
	\caption{ \label{aug3} Nuclear spin Rabi oscillations and the fitted Rabi frequencies and amplitudes as a function of magnetic field. (a) Nuclear Rabi oscillations for the off-axis transition $de$. An exponentially-damped sinusoid (solid line) is fitted to the experimental data (points). (b) Nuclear Rabi oscillations for transition $de$ for the off-axis orientations (top) and on-axis (bottom) for increasing magnetic field strength. The fitted normalized Rabi frequencies and amplitudes (inset) for these oscillations are shown in (c) together with the calculated augmentation factor for these transitions. (d) The normalized Rabi frequencies and amplitudes (inset) for transition $ab$ with the corresponding theoretical augmentation curve. The error bars in the data are the standard error in the fitted damped sinusoidal frequencies and amplitudes.}
	\vspace{-0.3cm}
\end{figure}

We then characterized the coupling between the P1 nuclear spin and the external rf field by measuring the P1 nuclear spin Rabi frequency as a function of magnetic field strength. Varying the length of the resonant rf pulse reveals Rabi oscillations of the nuclear spin (Fig. \ref{aug3}(a)). The spin-echo time was fixed at $^{13}$C revivals occurring at 45-65$\upmu$s to maintain consistent sensitivity between measurements as the magnetc field strength is changed. Using the labeling convention from Fig. \ref{deers2}(a), we identify two nuclear spin transitions, $\ket{a} \leftrightarrow \ket{b}$ and $\ket{d} \leftrightarrow \ket{e}$, which we denote as $ab$ and $de$. These particular transitions are well isolated from other resonant frequencies in the system across a range of magnetic fields and can therefore be individually addressed.

Figure \ref{aug3} (b) shows the measured Rabi oscillations for the $ab$ and $de$ transitions for magnetic field strengths between 10\,G and 100\,G. The contrast obtained for the off-axis measurements is approximately three times greater than for on-axis, due to the greater population of off-axis classes. For both the off-axis and on-axis orientations, the Rabi frequency and amplitude of the oscillations can be seen to decrease with increasing field strength. At each magnetic field, the resonant frequency to drive each transition is extracted from precise frequency-domain DEER measurements using low rf power, ensuring that the observed changes in the amplitude and frequency are not caused by off-resonant driving. The orientation of the rf field with respect to the P1 axes has only a minor effect on the gyromagnetic augmentation~\footnote{See Supplemental information}.

We fit a damped sinusoidal function of the form $S(t) = S(0) e^{(-t/T_D)^n} \cos{(\Omega t /2)}^2$ to the nuclear Rabi oscillation data, where $T_D$ is the effective decay time, to extract the Rabi frequency and oscillation amplitude for a given transition and magnetic field. The Rabi frequency is $\Omega = \alpha \,\gamma_N B_\text{rf}$, so for a constant rf field, the Rabi frequency scales with the effective gyromagnetic ratio of the nuclear spin as a function of the external magnetic field. The amplitude of the oscillation is determined by the coupling between the NV ensemble and the varying magnetic field produced by the nuclear spins. The magnetic moment of the nuclear spins is proportional to the effective gyromagnetic ratio and, therefore, the oscillation amplitude also scales with the augmentation factor.

The fitted frequencies and amplitudes of the Rabi oscillations for transition $de$ are presented in Fig. \ref{aug3}(c,e), together with the normalized augmentation factor calculated from Eq. \ref{alpha}. In order to effectively compare the results across different P1 transitions, both the Rabi frequencies and amplitudes were normalized with respect to the maximum values, with $\Omega_\text{max} = 1.82$ MHz and $S(0)_\text{max} = 0.012$ observed at 20\,G. The off-axis and on-axis data were then scaled according to the fitted frequencies and amplitudes measured for each transition at 35\,G. There is excellent agreement between the measured Rabi frequencies and the theoretical values for the augmentation factor of the nuclear spin gyromagnetic ratio. This agreement demonstrates that the augmented gyromagnetic ratio of the $^{14}$N nuclear spins is well explained by the hyperfine mixing with the P1 electron spin, which decreases with increasing magnetic field. 

The analysis of the nuclear Rabi oscillations for transition $ab$ (Fig. \ref{aug3}(d,f)) also shows an overall decrease in the frequency and amplitude with increasing magnetic field, which is well modeled by the augmentation curve for this transition. Given the low amplitude for the off-axis measurements for transition $ab$, it was not possible to obtain statistically significant measurements for the on-axis orientation for this particular transition. Furthermore, certain magnetic field values were excluded from the analysis for each transition where the resonant frequency at that field strength was too close to other transition frequencies to individually resolve. The results from our previous DEER measurements (Fig. \ref{deers2}) indicated that the nuclear spin features disappear from the observed signal as the magnetic field is increased. This observation is confirmed by the measured Rabi oscillation amplitude, which decrease markedly as the magnetic field is increased from 10\,G to 100\,G.


We have shown that the gyromagnetic ratio  of the P1 nuclear spin is augmented by  hyperfine coupling to the electron spin, and that the augmentation of the nuclear gyromagnetic ratio can be tuned by the external magnetic field. At low magnetic fields; that is, $<100$G, the augmentation factor increases sharply and we are able to perform rapid quantum control of the nuclear spin state, at Rabi frequencies comparable to the P1 or NV electron spin.   

A remaining open question, the subject of further work, concerns the consequences of gyromagnetic augmentation on the coherence of the P1 nuclear spins. As more electron spin character is mixed into the nuclear spin, the coupling to not only rf fields increases (as studied in this work) but also to magnetic field noise which suppresses spin coherence. However, there is reason to expect the augmented nuclear spins to still possess coherence times significantly exceeding that of the P1 or NV electron spin if not the bare nuclear spin coherence time. The NV nuclear spin augmentation factor is around $\alpha = 20$ at 500\,G, but the response to static magnetic fields is still set by the bare value of $\gamma_N$. As recently reported in Ref~\cite{degen_entanglement_2021}, augmented nuclear spin transitions at low field possess coherence times a factor of 4 greater than that of the P1 electron spin. Furthermore, rapid state control and long spin storage can potentially be combined with high speed magnetic shuttling to dynamically tune the nuclear spin interaction with control fields. In this work we examined a thermally distributed ensemble of P1 spins, but with augmented driving, rapid hyperpolarisation of the P1 nuclear spins may allow substantial enhancement, which is a possible avenue of future study.

Our work also shows explicitly that the NV center is sensitive to the P1 nuclear spin state at weak magnetic fields, adding further channels for decoherence in NV quantum sensing. The detection of nuclear Rabi oscillations in the NV spin-echo signal implies that the P1 nuclear spins are a significant component in the interacting environment of the NV and will contribute to the dephasing of NV spins at low magnetic fields. Suppression of NV dephasing through spin-bath control and dynamical decoupling schemes \cite{de_lange_controlling_2012, bauch_ultralong_2018} could, therefore, be enhanced at lower magnetic fields by addressing the nuclear spins in addition to the P1 electron spins. At higher densities, the dipolar coupling between the NV and P1 electrons may be used to indirectly couple the associated nuclear spins, suggesting an interesting platform for quantum information processing~\cite{childress_diamond_2013}.

In conclusion, we have characterized the magnetic-field induced augmentation of the gyromagnetic ratio for the P1 nuclear spin. We have found that at lower magnetic field magnitudes, we are able to enact fast quantum control of the nuclear spins within the defect. Tunable magnetic augmentation establishes a potential path forward for using the P1 nuclear spin as a qubit in quantum applications, by addressing the challenge of accessing and controlling nuclear spins that are typically well isolated from external fields.

\section*{Acknowledgements}
\vspace{-0.3cm} We thank R. E. Scholten for a careful reading of the manuscript and constructive discussions. This work was supported by the Australian Research Council Discovery Scheme (DP190100949). R. M. G. was supported by a Research Training Program Scholarship.

\bibliographystyle{apsrev4-2}

\begin{thebibliography}{49}%
\makeatletter
\providecommand \@ifxundefined [1]{%
 \@ifx{#1\undefined}
}%
\providecommand \@ifnum [1]{%
 \ifnum #1\expandafter \@firstoftwo
 \else \expandafter \@secondoftwo
 \fi
}%
\providecommand \@ifx [1]{%
 \ifx #1\expandafter \@firstoftwo
 \else \expandafter \@secondoftwo
 \fi
}%
\providecommand \natexlab [1]{#1}%
\providecommand \enquote  [1]{``#1''}%
\providecommand \bibnamefont  [1]{#1}%
\providecommand \bibfnamefont [1]{#1}%
\providecommand \citenamefont [1]{#1}%
\providecommand \href@noop [0]{\@secondoftwo}%
\providecommand \href [0]{\begingroup \@sanitize@url \@href}%
\providecommand \@href[1]{\@@startlink{#1}\@@href}%
\providecommand \@@href[1]{\endgroup#1\@@endlink}%
\providecommand \@sanitize@url [0]{\catcode `\\12\catcode `\$12\catcode
  `\&12\catcode `\#12\catcode `\^12\catcode `\_12\catcode `\%12\relax}%
\providecommand \@@startlink[1]{}%
\providecommand \@@endlink[0]{}%
\providecommand \url  [0]{\begingroup\@sanitize@url \@url }%
\providecommand \@url [1]{\endgroup\@href {#1}{\urlprefix }}%
\providecommand \urlprefix  [0]{URL }%
\providecommand \Eprint [0]{\href }%
\providecommand \doibase [0]{https://doi.org/}%
\providecommand \selectlanguage [0]{\@gobble}%
\providecommand \bibinfo  [0]{\@secondoftwo}%
\providecommand \bibfield  [0]{\@secondoftwo}%
\providecommand \translation [1]{[#1]}%
\providecommand \BibitemOpen [0]{}%
\providecommand \bibitemStop [0]{}%
\providecommand \bibitemNoStop [0]{.\EOS\space}%
\providecommand \EOS [0]{\spacefactor3000\relax}%
\providecommand \BibitemShut  [1]{\csname bibitem#1\endcsname}%
\let\auto@bib@innerbib\@empty
\bibitem [{\citenamefont {Fuchs}\ \emph {et~al.}(2011)\citenamefont {Fuchs},
  \citenamefont {Burkard}, \citenamefont {Klimov},\ and\ \citenamefont
  {Awschalom}}]{fuchs_quantum_2011}%
  \BibitemOpen
  \bibfield  {author} {\bibinfo {author} {\bibfnamefont {G.~D.}\ \bibnamefont
  {Fuchs}}, \bibinfo {author} {\bibfnamefont {G.}~\bibnamefont {Burkard}},
  \bibinfo {author} {\bibfnamefont {P.~V.}\ \bibnamefont {Klimov}},\ and\
  \bibinfo {author} {\bibfnamefont {D.~D.}\ \bibnamefont {Awschalom}},\
  }\bibfield  {title} {{\selectlanguage {en}\bibinfo {title} {A quantum memory
  intrinsic to single nitrogen–vacancy centres in diamond}},\ }\bibfield
  {journal} {\bibinfo  {journal} {Nature Phys}\ }\textbf {\bibinfo {volume}
  {7}},\ \href {https://doi.org/10.1038/nphys2026} {10.1038/nphys2026}
  (\bibinfo {year} {2011})\BibitemShut {NoStop}%
\bibitem [{\citenamefont {Pla}\ \emph {et~al.}(2014)\citenamefont {Pla},
  \citenamefont {Mohiyaddin}, \citenamefont {Tan}, \citenamefont {Dehollain},
  \citenamefont {Rahman}, \citenamefont {Klimeck}, \citenamefont {Jamieson},
  \citenamefont {Dzurak},\ and\ \citenamefont {Morello}}]{pla_coherent_2014}%
  \BibitemOpen
  \bibfield  {author} {\bibinfo {author} {\bibfnamefont {J.~J.}\ \bibnamefont
  {Pla}}, \bibinfo {author} {\bibfnamefont {F.~A.}\ \bibnamefont {Mohiyaddin}},
  \bibinfo {author} {\bibfnamefont {K.~Y.}\ \bibnamefont {Tan}}, \bibinfo
  {author} {\bibfnamefont {J.~P.}\ \bibnamefont {Dehollain}}, \bibinfo {author}
  {\bibfnamefont {R.}~\bibnamefont {Rahman}}, \bibinfo {author} {\bibfnamefont
  {G.}~\bibnamefont {Klimeck}}, \bibinfo {author} {\bibfnamefont {D.~N.}\
  \bibnamefont {Jamieson}}, \bibinfo {author} {\bibfnamefont {A.~S.}\
  \bibnamefont {Dzurak}},\ and\ \bibinfo {author} {\bibfnamefont
  {A.}~\bibnamefont {Morello}},\ }\bibfield  {title} {\bibinfo {title}
  {Coherent {Control} of a {Single} {$^{29}\mathrm{Si}$} {Nuclear} {Spin}
  {Qubit}},\ }\href {https://doi.org/10.1103/PhysRevLett.113.246801} {\bibfield
   {journal} {\bibinfo  {journal} {Phys. Rev. Lett.}\ }\textbf {\bibinfo
  {volume} {113}},\ \bibinfo {pages} {246801} (\bibinfo {year}
  {2014})}\BibitemShut {NoStop}%
\bibitem [{\citenamefont {Hensen}\ \emph {et~al.}(2020)\citenamefont {Hensen},
  \citenamefont {Wei~Huang}, \citenamefont {Yang}, \citenamefont {Wai~Chan},
  \citenamefont {Yoneda}, \citenamefont {Tanttu}, \citenamefont {Hudson},
  \citenamefont {Laucht}, \citenamefont {Itoh}, \citenamefont {Ladd},
  \citenamefont {Morello},\ and\ \citenamefont {Dzurak}}]{hensen_silicon_2020}%
  \BibitemOpen
  \bibfield  {author} {\bibinfo {author} {\bibfnamefont {B.}~\bibnamefont
  {Hensen}}, \bibinfo {author} {\bibfnamefont {W.}~\bibnamefont {Wei~Huang}},
  \bibinfo {author} {\bibfnamefont {C.-H.}\ \bibnamefont {Yang}}, \bibinfo
  {author} {\bibfnamefont {K.}~\bibnamefont {Wai~Chan}}, \bibinfo {author}
  {\bibfnamefont {J.}~\bibnamefont {Yoneda}}, \bibinfo {author} {\bibfnamefont
  {T.}~\bibnamefont {Tanttu}}, \bibinfo {author} {\bibfnamefont {F.~E.}\
  \bibnamefont {Hudson}}, \bibinfo {author} {\bibfnamefont {A.}~\bibnamefont
  {Laucht}}, \bibinfo {author} {\bibfnamefont {K.~M.}\ \bibnamefont {Itoh}},
  \bibinfo {author} {\bibfnamefont {T.~D.}\ \bibnamefont {Ladd}}, \bibinfo
  {author} {\bibfnamefont {A.}~\bibnamefont {Morello}},\ and\ \bibinfo {author}
  {\bibfnamefont {A.~S.}\ \bibnamefont {Dzurak}},\ }\bibfield  {title}
  {{\selectlanguage {en}\bibinfo {title} {A silicon quantum-dot-coupled nuclear
  spin qubit}},\ }\href {https://doi.org/10.1038/s41565-019-0587-7} {\bibfield
  {journal} {\bibinfo  {journal} {Nat. Nanotechnol.}\ }\textbf {\bibinfo
  {volume} {15}},\ \bibinfo {pages} {13} (\bibinfo {year} {2020})}\BibitemShut
  {NoStop}%
\bibitem [{\citenamefont {Muhonen}\ \emph {et~al.}(2014)\citenamefont
  {Muhonen}, \citenamefont {Dehollain}, \citenamefont {Laucht}, \citenamefont
  {Hudson}, \citenamefont {Kalra}, \citenamefont {Sekiguchi}, \citenamefont
  {Itoh}, \citenamefont {Jamieson}, \citenamefont {McCallum}, \citenamefont
  {Dzurak},\ and\ \citenamefont {Morello}}]{muhonen_storing_2014}%
  \BibitemOpen
  \bibfield  {author} {\bibinfo {author} {\bibfnamefont {J.~T.}\ \bibnamefont
  {Muhonen}}, \bibinfo {author} {\bibfnamefont {J.~P.}\ \bibnamefont
  {Dehollain}}, \bibinfo {author} {\bibfnamefont {A.}~\bibnamefont {Laucht}},
  \bibinfo {author} {\bibfnamefont {F.~E.}\ \bibnamefont {Hudson}}, \bibinfo
  {author} {\bibfnamefont {R.}~\bibnamefont {Kalra}}, \bibinfo {author}
  {\bibfnamefont {T.}~\bibnamefont {Sekiguchi}}, \bibinfo {author}
  {\bibfnamefont {K.~M.}\ \bibnamefont {Itoh}}, \bibinfo {author}
  {\bibfnamefont {D.~N.}\ \bibnamefont {Jamieson}}, \bibinfo {author}
  {\bibfnamefont {J.~C.}\ \bibnamefont {McCallum}}, \bibinfo {author}
  {\bibfnamefont {A.~S.}\ \bibnamefont {Dzurak}},\ and\ \bibinfo {author}
  {\bibfnamefont {A.}~\bibnamefont {Morello}},\ }\bibfield  {title}
  {{\selectlanguage {en}\bibinfo {title} {Storing quantum information for 30
  seconds in a nanoelectronic device}},\ }\href
  {https://doi.org/10.1038/nnano.2014.211} {\bibfield  {journal} {\bibinfo
  {journal} {Nature Nanotech}\ }\textbf {\bibinfo {volume} {9}},\ \bibinfo
  {pages} {986} (\bibinfo {year} {2014})}\BibitemShut {NoStop}%
\bibitem [{\citenamefont {Degen}\ \emph {et~al.}(2017)\citenamefont {Degen},
  \citenamefont {Reinhard},\ and\ \citenamefont
  {Cappellaro}}]{degen_quantum_2017}%
  \BibitemOpen
  \bibfield  {author} {\bibinfo {author} {\bibfnamefont {C.}~\bibnamefont
  {Degen}}, \bibinfo {author} {\bibfnamefont {F.}~\bibnamefont {Reinhard}},\
  and\ \bibinfo {author} {\bibfnamefont {P.}~\bibnamefont {Cappellaro}},\
  }\bibfield  {title} {\bibinfo {title} {Quantum sensing},\ }\href
  {https://doi.org/10.1103/RevModPhys.89.035002} {\bibfield  {journal}
  {\bibinfo  {journal} {Rev. Mod. Phys.}\ }\textbf {\bibinfo {volume} {89}},\
  \bibinfo {pages} {035002} (\bibinfo {year} {2017})}\BibitemShut {NoStop}%
\bibitem [{\citenamefont {Maclaurin}\ \emph {et~al.}(2012)\citenamefont
  {Maclaurin}, \citenamefont {Doherty}, \citenamefont {Hollenberg},\ and\
  \citenamefont {Martin}}]{maclaurin_measurable_2012}%
  \BibitemOpen
  \bibfield  {author} {\bibinfo {author} {\bibfnamefont {D.}~\bibnamefont
  {Maclaurin}}, \bibinfo {author} {\bibfnamefont {M.~W.}\ \bibnamefont
  {Doherty}}, \bibinfo {author} {\bibfnamefont {L.~C.~L.}\ \bibnamefont
  {Hollenberg}},\ and\ \bibinfo {author} {\bibfnamefont {A.~M.}\ \bibnamefont
  {Martin}},\ }\bibfield  {title} {\bibinfo {title} {Measurable {Quantum}
  {Geometric} {Phase} from a {Rotating} {Single} {Spin}},\ }\href
  {https://doi.org/10.1103/PhysRevLett.108.240403} {\bibfield  {journal}
  {\bibinfo  {journal} {Phys. Rev. Lett.}\ }\textbf {\bibinfo {volume} {108}},\
  \bibinfo {pages} {240403} (\bibinfo {year} {2012})}\BibitemShut {NoStop}%
\bibitem [{\citenamefont {Ledbetter}\ \emph {et~al.}(2012)\citenamefont
  {Ledbetter}, \citenamefont {Jensen}, \citenamefont {Fischer}, \citenamefont
  {Jarmola},\ and\ \citenamefont {Budker}}]{ledbetter_gyroscopes_2012}%
  \BibitemOpen
  \bibfield  {author} {\bibinfo {author} {\bibfnamefont {M.~P.}\ \bibnamefont
  {Ledbetter}}, \bibinfo {author} {\bibfnamefont {K.}~\bibnamefont {Jensen}},
  \bibinfo {author} {\bibfnamefont {R.}~\bibnamefont {Fischer}}, \bibinfo
  {author} {\bibfnamefont {A.}~\bibnamefont {Jarmola}},\ and\ \bibinfo {author}
  {\bibfnamefont {D.}~\bibnamefont {Budker}},\ }\bibfield  {title} {\bibinfo
  {title} {Gyroscopes based on nitrogen-vacancy centers in diamond},\ }\href
  {https://doi.org/10.1103/PhysRevA.86.052116} {\bibfield  {journal} {\bibinfo
  {journal} {Phys. Rev. A}\ }\textbf {\bibinfo {volume} {86}},\ \bibinfo
  {pages} {052116} (\bibinfo {year} {2012})}\BibitemShut {NoStop}%
\bibitem [{\citenamefont {Ajoy}\ and\ \citenamefont
  {Cappellaro}(2012)}]{ajoy_stable_2012}%
  \BibitemOpen
  \bibfield  {author} {\bibinfo {author} {\bibfnamefont {A.}~\bibnamefont
  {Ajoy}}\ and\ \bibinfo {author} {\bibfnamefont {P.}~\bibnamefont
  {Cappellaro}},\ }\bibfield  {title} {\bibinfo {title} {Stable three-axis
  nuclear-spin gyroscope in diamond},\ }\href
  {https://doi.org/10.1103/PhysRevA.86.062104} {\bibfield  {journal} {\bibinfo
  {journal} {Phys. Rev. A}\ }\textbf {\bibinfo {volume} {86}},\ \bibinfo
  {pages} {062104} (\bibinfo {year} {2012})}\BibitemShut {NoStop}%
\bibitem [{\citenamefont {Liu}\ \emph {et~al.}(2019)\citenamefont {Liu},
  \citenamefont {Ajoy},\ and\ \citenamefont {Cappellaro}}]{liu_nanoscale_2019}%
  \BibitemOpen
  \bibfield  {author} {\bibinfo {author} {\bibfnamefont {Y.-X.}\ \bibnamefont
  {Liu}}, \bibinfo {author} {\bibfnamefont {A.}~\bibnamefont {Ajoy}},\ and\
  \bibinfo {author} {\bibfnamefont {P.}~\bibnamefont {Cappellaro}},\ }\bibfield
   {title} {\bibinfo {title} {Nanoscale {Vector} dc {Magnetometry} via
  {Ancilla}-{Assisted} {Frequency} {Up}-{Conversion}},\ }\href
  {https://doi.org/10.1103/PhysRevLett.122.100501} {\bibfield  {journal}
  {\bibinfo  {journal} {Phys. Rev. Lett.}\ }\textbf {\bibinfo {volume} {122}},\
  \bibinfo {pages} {100501} (\bibinfo {year} {2019})}\BibitemShut {NoStop}%
\bibitem [{\citenamefont {Soshenko}\ \emph {et~al.}(2021)\citenamefont
  {Soshenko}, \citenamefont {Bolshedvorskii}, \citenamefont {Rubinas},
  \citenamefont {Sorokin}, \citenamefont {Smolyaninov}, \citenamefont
  {Vorobyov},\ and\ \citenamefont {Akimov}}]{soshenko_nuclear_2021}%
  \BibitemOpen
  \bibfield  {author} {\bibinfo {author} {\bibfnamefont {V.~V.}\ \bibnamefont
  {Soshenko}}, \bibinfo {author} {\bibfnamefont {S.~V.}\ \bibnamefont
  {Bolshedvorskii}}, \bibinfo {author} {\bibfnamefont {O.}~\bibnamefont
  {Rubinas}}, \bibinfo {author} {\bibfnamefont {V.~N.}\ \bibnamefont
  {Sorokin}}, \bibinfo {author} {\bibfnamefont {A.~N.}\ \bibnamefont
  {Smolyaninov}}, \bibinfo {author} {\bibfnamefont {V.~V.}\ \bibnamefont
  {Vorobyov}},\ and\ \bibinfo {author} {\bibfnamefont {A.~V.}\ \bibnamefont
  {Akimov}},\ }\bibfield  {title} {\bibinfo {title} {Nuclear {Spin} {Gyroscope}
  based on the {Nitrogen} {Vacancy} {Center} in {Diamond}},\ }\href
  {https://doi.org/10.1103/PhysRevLett.126.197702} {\bibfield  {journal}
  {\bibinfo  {journal} {Phys. Rev. Lett.}\ }\textbf {\bibinfo {volume} {126}},\
  \bibinfo {pages} {197702} (\bibinfo {year} {2021})}\BibitemShut {NoStop}%
\bibitem [{\citenamefont {Jarmola}\ \emph {et~al.}(2021)\citenamefont
  {Jarmola}, \citenamefont {Lourette}, \citenamefont {Acosta}, \citenamefont
  {Birdwell}, \citenamefont {Blümler}, \citenamefont {Budker}, \citenamefont
  {Ivanov},\ and\ \citenamefont {Malinovsky}}]{jarmola_demonstration_2021}%
  \BibitemOpen
  \bibfield  {author} {\bibinfo {author} {\bibfnamefont {A.}~\bibnamefont
  {Jarmola}}, \bibinfo {author} {\bibfnamefont {S.}~\bibnamefont {Lourette}},
  \bibinfo {author} {\bibfnamefont {V.~M.}\ \bibnamefont {Acosta}}, \bibinfo
  {author} {\bibfnamefont {A.~G.}\ \bibnamefont {Birdwell}}, \bibinfo {author}
  {\bibfnamefont {P.}~\bibnamefont {Blümler}}, \bibinfo {author}
  {\bibfnamefont {D.}~\bibnamefont {Budker}}, \bibinfo {author} {\bibfnamefont
  {T.}~\bibnamefont {Ivanov}},\ and\ \bibinfo {author} {\bibfnamefont {V.~S.}\
  \bibnamefont {Malinovsky}},\ }\bibfield  {title} {{\selectlanguage
  {en}\bibinfo {title} {Demonstration of diamond nuclear spin gyroscope}},\
  }\href {https://arxiv.org/abs/2107.04257v1} {\  (\bibinfo {year}
  {2021})}\BibitemShut {NoStop}%
\bibitem [{\citenamefont {Kane}(1998)}]{kane_silicon-based_1998}%
  \BibitemOpen
  \bibfield  {author} {\bibinfo {author} {\bibfnamefont {B.~E.}\ \bibnamefont
  {Kane}},\ }\bibfield  {title} {{\selectlanguage {en}\bibinfo {title} {A
  silicon-based nuclear spin quantum computer}},\ }\href
  {https://doi.org/10.1038/30156} {\bibfield  {journal} {\bibinfo  {journal}
  {Nature}\ }\textbf {\bibinfo {volume} {393}},\ \bibinfo {pages} {133}
  (\bibinfo {year} {1998})}\BibitemShut {NoStop}%
\bibitem [{\citenamefont {Dutt}\ \emph {et~al.}(2007)\citenamefont {Dutt},
  \citenamefont {Childress}, \citenamefont {Jiang}, \citenamefont {Togan},
  \citenamefont {Maze}, \citenamefont {Jelezko}, \citenamefont {Zibrov},
  \citenamefont {Hemmer},\ and\ \citenamefont {Lukin}}]{dutt_quantum_2007}%
  \BibitemOpen
  \bibfield  {author} {\bibinfo {author} {\bibfnamefont {M.~V.~G.}\
  \bibnamefont {Dutt}}, \bibinfo {author} {\bibfnamefont {L.}~\bibnamefont
  {Childress}}, \bibinfo {author} {\bibfnamefont {L.}~\bibnamefont {Jiang}},
  \bibinfo {author} {\bibfnamefont {E.}~\bibnamefont {Togan}}, \bibinfo
  {author} {\bibfnamefont {J.}~\bibnamefont {Maze}}, \bibinfo {author}
  {\bibfnamefont {F.}~\bibnamefont {Jelezko}}, \bibinfo {author} {\bibfnamefont
  {A.~S.}\ \bibnamefont {Zibrov}}, \bibinfo {author} {\bibfnamefont {P.~R.}\
  \bibnamefont {Hemmer}},\ and\ \bibinfo {author} {\bibfnamefont {M.~D.}\
  \bibnamefont {Lukin}},\ }\bibfield  {title} {\bibinfo {title} {Quantum
  {Register} {Based} on {Individual} {Electronic} and {Nuclear} {Spin} {Qubits}
  in {Diamond}},\ }\href {https://doi.org/10.1126/science.1139831} {\bibfield
  {journal} {\bibinfo  {journal} {Science}\ }\textbf {\bibinfo {volume}
  {316}},\ \bibinfo {pages} {1312} (\bibinfo {year} {2007})}\BibitemShut
  {NoStop}%
\bibitem [{\citenamefont {Ladd}\ \emph {et~al.}(2010)\citenamefont {Ladd},
  \citenamefont {Jelezko}, \citenamefont {Laflamme}, \citenamefont {Nakamura},
  \citenamefont {Monroe},\ and\ \citenamefont {O’Brien}}]{ladd_quantum_2010}%
  \BibitemOpen
  \bibfield  {author} {\bibinfo {author} {\bibfnamefont {T.~D.}\ \bibnamefont
  {Ladd}}, \bibinfo {author} {\bibfnamefont {F.}~\bibnamefont {Jelezko}},
  \bibinfo {author} {\bibfnamefont {R.}~\bibnamefont {Laflamme}}, \bibinfo
  {author} {\bibfnamefont {Y.}~\bibnamefont {Nakamura}}, \bibinfo {author}
  {\bibfnamefont {C.}~\bibnamefont {Monroe}},\ and\ \bibinfo {author}
  {\bibfnamefont {J.~L.}\ \bibnamefont {O’Brien}},\ }\bibfield  {title}
  {{\selectlanguage {en}\bibinfo {title} {Quantum computers}},\ }\href
  {https://doi.org/10.1038/nature08812} {\bibfield  {journal} {\bibinfo
  {journal} {Nature}\ }\textbf {\bibinfo {volume} {464}},\ \bibinfo {pages}
  {45} (\bibinfo {year} {2010})}\BibitemShut {NoStop}%
\bibitem [{\citenamefont {Maurer}\ \emph {et~al.}(2012)\citenamefont {Maurer},
  \citenamefont {Kucsko}, \citenamefont {Latta}, \citenamefont {Jiang},
  \citenamefont {Yao}, \citenamefont {Bennett}, \citenamefont {Pastawski},
  \citenamefont {Hunger}, \citenamefont {Chisholm}, \citenamefont {Markham},
  \citenamefont {Twitchen}, \citenamefont {Cirac},\ and\ \citenamefont
  {Lukin}}]{maurer_room-temperature_2012}%
  \BibitemOpen
  \bibfield  {author} {\bibinfo {author} {\bibfnamefont {P.~C.}\ \bibnamefont
  {Maurer}}, \bibinfo {author} {\bibfnamefont {G.}~\bibnamefont {Kucsko}},
  \bibinfo {author} {\bibfnamefont {C.}~\bibnamefont {Latta}}, \bibinfo
  {author} {\bibfnamefont {L.}~\bibnamefont {Jiang}}, \bibinfo {author}
  {\bibfnamefont {N.~Y.}\ \bibnamefont {Yao}}, \bibinfo {author} {\bibfnamefont
  {S.~D.}\ \bibnamefont {Bennett}}, \bibinfo {author} {\bibfnamefont
  {F.}~\bibnamefont {Pastawski}}, \bibinfo {author} {\bibfnamefont
  {D.}~\bibnamefont {Hunger}}, \bibinfo {author} {\bibfnamefont
  {N.}~\bibnamefont {Chisholm}}, \bibinfo {author} {\bibfnamefont
  {M.}~\bibnamefont {Markham}}, \bibinfo {author} {\bibfnamefont {D.~J.}\
  \bibnamefont {Twitchen}}, \bibinfo {author} {\bibfnamefont {J.~I.}\
  \bibnamefont {Cirac}},\ and\ \bibinfo {author} {\bibfnamefont {M.~D.}\
  \bibnamefont {Lukin}},\ }\bibfield  {title} {\bibinfo {title}
  {Room-{Temperature} {Quantum} {Bit} {Memory} {Exceeding} {One} {Second}},\
  }\href {https://doi.org/10.1126/science.1220513} {\bibfield  {journal}
  {\bibinfo  {journal} {Science}\ }\textbf {\bibinfo {volume} {336}},\ \bibinfo
  {pages} {1283} (\bibinfo {year} {2012})}\BibitemShut {NoStop}%
\bibitem [{\citenamefont {Zhong}\ \emph {et~al.}(2015)\citenamefont {Zhong},
  \citenamefont {Hedges}, \citenamefont {Ahlefeldt}, \citenamefont
  {Bartholomew}, \citenamefont {Beavan}, \citenamefont {Wittig}, \citenamefont
  {Longdell},\ and\ \citenamefont {Sellars}}]{zhong_optically_2015}%
  \BibitemOpen
  \bibfield  {author} {\bibinfo {author} {\bibfnamefont {M.}~\bibnamefont
  {Zhong}}, \bibinfo {author} {\bibfnamefont {M.~P.}\ \bibnamefont {Hedges}},
  \bibinfo {author} {\bibfnamefont {R.~L.}\ \bibnamefont {Ahlefeldt}}, \bibinfo
  {author} {\bibfnamefont {J.~G.}\ \bibnamefont {Bartholomew}}, \bibinfo
  {author} {\bibfnamefont {S.~E.}\ \bibnamefont {Beavan}}, \bibinfo {author}
  {\bibfnamefont {S.~M.}\ \bibnamefont {Wittig}}, \bibinfo {author}
  {\bibfnamefont {J.~J.}\ \bibnamefont {Longdell}},\ and\ \bibinfo {author}
  {\bibfnamefont {M.~J.}\ \bibnamefont {Sellars}},\ }\bibfield  {title}
  {{\selectlanguage {en}\bibinfo {title} {Optically addressable nuclear spins
  in a solid with a six-hour coherence time}},\ }\href
  {https://doi.org/10.1038/nature14025} {\bibfield  {journal} {\bibinfo
  {journal} {Nature}\ }\textbf {\bibinfo {volume} {517}},\ \bibinfo {pages}
  {177} (\bibinfo {year} {2015})}\BibitemShut {NoStop}%
\bibitem [{\citenamefont {Doherty}\ \emph {et~al.}(2013)\citenamefont
  {Doherty}, \citenamefont {Manson}, \citenamefont {Delaney}, \citenamefont
  {Jelezko}, \citenamefont {Wrachtrup},\ and\ \citenamefont
  {Hollenberg}}]{doherty_nitrogen-vacancy_2013}%
  \BibitemOpen
  \bibfield  {author} {\bibinfo {author} {\bibfnamefont {M.~W.}\ \bibnamefont
  {Doherty}}, \bibinfo {author} {\bibfnamefont {N.~B.}\ \bibnamefont {Manson}},
  \bibinfo {author} {\bibfnamefont {P.}~\bibnamefont {Delaney}}, \bibinfo
  {author} {\bibfnamefont {F.}~\bibnamefont {Jelezko}}, \bibinfo {author}
  {\bibfnamefont {J.}~\bibnamefont {Wrachtrup}},\ and\ \bibinfo {author}
  {\bibfnamefont {L.~C.~L.}\ \bibnamefont {Hollenberg}},\ }\bibfield  {title}
  {{\selectlanguage {en}\bibinfo {title} {The nitrogen-vacancy colour centre in
  diamond}},\ }\href {https://doi.org/10.1016/j.physrep.2013.02.001} {\bibfield
   {journal} {\bibinfo  {journal} {Physics Reports}\ }\bibinfo {series} {The
  nitrogen-vacancy colour centre in diamond},\ \textbf {\bibinfo {volume}
  {528}},\ \bibinfo {pages} {1} (\bibinfo {year} {2013})}\BibitemShut {NoStop}%
\bibitem [{\citenamefont {Schirhagl}\ \emph {et~al.}(2014)\citenamefont
  {Schirhagl}, \citenamefont {Chang}, \citenamefont {Loretz},\ and\
  \citenamefont {Degen}}]{schirhagl_nitrogen-vacancy_2014}%
  \BibitemOpen
  \bibfield  {author} {\bibinfo {author} {\bibfnamefont {R.}~\bibnamefont
  {Schirhagl}}, \bibinfo {author} {\bibfnamefont {K.}~\bibnamefont {Chang}},
  \bibinfo {author} {\bibfnamefont {M.}~\bibnamefont {Loretz}},\ and\ \bibinfo
  {author} {\bibfnamefont {C.~L.}\ \bibnamefont {Degen}},\ }\bibfield  {title}
  {\bibinfo {title} {Nitrogen-{Vacancy} {Centers} in {Diamond}: {Nanoscale}
  {Sensors} for {Physics} and {Biology}},\ }\href
  {https://doi.org/10.1146/annurev-physchem-040513-103659} {\bibfield
  {journal} {\bibinfo  {journal} {Annual Review of Physical Chemistry}\
  }\textbf {\bibinfo {volume} {65}},\ \bibinfo {pages} {83} (\bibinfo {year}
  {2014})}\BibitemShut {NoStop}%
\bibitem [{\citenamefont {Wu}\ \emph {et~al.}(2016)\citenamefont {Wu},
  \citenamefont {Jelezko}, \citenamefont {Plenio},\ and\ \citenamefont
  {Weil}}]{wu_diamond_2016}%
  \BibitemOpen
  \bibfield  {author} {\bibinfo {author} {\bibfnamefont {Y.}~\bibnamefont
  {Wu}}, \bibinfo {author} {\bibfnamefont {F.}~\bibnamefont {Jelezko}},
  \bibinfo {author} {\bibfnamefont {M.~B.}\ \bibnamefont {Plenio}},\ and\
  \bibinfo {author} {\bibfnamefont {T.}~\bibnamefont {Weil}},\ }\bibfield
  {title} {{\selectlanguage {eng}\bibinfo {title} {Diamond {Quantum} {Devices}
  in {Biology}}},\ }\href {https://doi.org/10.1002/anie.201506556} {\bibfield
  {journal} {\bibinfo  {journal} {Angew Chem Int Ed Engl}\ }\textbf {\bibinfo
  {volume} {55}},\ \bibinfo {pages} {6586} (\bibinfo {year}
  {2016})}\BibitemShut {NoStop}%
\bibitem [{\citenamefont {Jacques}\ \emph {et~al.}(2009)\citenamefont
  {Jacques}, \citenamefont {Neumann}, \citenamefont {Beck}, \citenamefont
  {Markham}, \citenamefont {Twitchen}, \citenamefont {Meijer}, \citenamefont
  {Kaiser}, \citenamefont {Balasubramanian}, \citenamefont {Jelezko},\ and\
  \citenamefont {Wrachtrup}}]{jacques_dynamic_2009}%
  \BibitemOpen
  \bibfield  {author} {\bibinfo {author} {\bibfnamefont {V.}~\bibnamefont
  {Jacques}}, \bibinfo {author} {\bibfnamefont {P.}~\bibnamefont {Neumann}},
  \bibinfo {author} {\bibfnamefont {J.}~\bibnamefont {Beck}}, \bibinfo {author}
  {\bibfnamefont {M.}~\bibnamefont {Markham}}, \bibinfo {author} {\bibfnamefont
  {D.}~\bibnamefont {Twitchen}}, \bibinfo {author} {\bibfnamefont
  {J.}~\bibnamefont {Meijer}}, \bibinfo {author} {\bibfnamefont
  {F.}~\bibnamefont {Kaiser}}, \bibinfo {author} {\bibfnamefont
  {G.}~\bibnamefont {Balasubramanian}}, \bibinfo {author} {\bibfnamefont
  {F.}~\bibnamefont {Jelezko}},\ and\ \bibinfo {author} {\bibfnamefont
  {J.}~\bibnamefont {Wrachtrup}},\ }\bibfield  {title} {\bibinfo {title}
  {Dynamic {Polarization} of {Single} {Nuclear} {Spins} by {Optical} {Pumping}
  of {Nitrogen}-{Vacancy} {Color} {Centers} in {Diamond} at {Room}
  {Temperature}},\ }\href {https://doi.org/10.1103/PhysRevLett.102.057403}
  {\bibfield  {journal} {\bibinfo  {journal} {Phys. Rev. Lett.}\ }\textbf
  {\bibinfo {volume} {102}},\ \bibinfo {pages} {057403} (\bibinfo {year}
  {2009})}\BibitemShut {NoStop}%
\bibitem [{\citenamefont {Neumann}\ \emph {et~al.}(2010)\citenamefont
  {Neumann}, \citenamefont {Beck}, \citenamefont {Steiner}, \citenamefont
  {Rempp}, \citenamefont {Fedder}, \citenamefont {Hemmer}, \citenamefont
  {Wrachtrup},\ and\ \citenamefont {Jelezko}}]{neumann_single-shot_2010}%
  \BibitemOpen
  \bibfield  {author} {\bibinfo {author} {\bibfnamefont {P.}~\bibnamefont
  {Neumann}}, \bibinfo {author} {\bibfnamefont {J.}~\bibnamefont {Beck}},
  \bibinfo {author} {\bibfnamefont {M.}~\bibnamefont {Steiner}}, \bibinfo
  {author} {\bibfnamefont {F.}~\bibnamefont {Rempp}}, \bibinfo {author}
  {\bibfnamefont {H.}~\bibnamefont {Fedder}}, \bibinfo {author} {\bibfnamefont
  {P.~R.}\ \bibnamefont {Hemmer}}, \bibinfo {author} {\bibfnamefont
  {J.}~\bibnamefont {Wrachtrup}},\ and\ \bibinfo {author} {\bibfnamefont
  {F.}~\bibnamefont {Jelezko}},\ }\bibfield  {title} {\bibinfo {title}
  {Single-{Shot} {Readout} of a {Single} {Nuclear} {Spin}},\ }\href
  {https://doi.org/10.1126/science.1189075} {\bibfield  {journal} {\bibinfo
  {journal} {Science}\ }\textbf {\bibinfo {volume} {329}},\ \bibinfo {pages}
  {542} (\bibinfo {year} {2010})}\BibitemShut {NoStop}%
\bibitem [{\citenamefont {Smeltzer}\ \emph {et~al.}(2009)\citenamefont
  {Smeltzer}, \citenamefont {McIntyre},\ and\ \citenamefont
  {Childress}}]{smeltzer_robust_2009}%
  \BibitemOpen
  \bibfield  {author} {\bibinfo {author} {\bibfnamefont {B.}~\bibnamefont
  {Smeltzer}}, \bibinfo {author} {\bibfnamefont {J.}~\bibnamefont {McIntyre}},\
  and\ \bibinfo {author} {\bibfnamefont {L.}~\bibnamefont {Childress}},\
  }\bibfield  {title} {\bibinfo {title} {Robust control of individual nuclear
  spins in diamond},\ }\href {https://doi.org/10.1103/PhysRevA.80.050302}
  {\bibfield  {journal} {\bibinfo  {journal} {Phys. Rev. A}\ }\textbf {\bibinfo
  {volume} {80}},\ \bibinfo {pages} {050302} (\bibinfo {year}
  {2009})}\BibitemShut {NoStop}%
\bibitem [{\citenamefont {Sangtawesin}\ \emph {et~al.}(2016)\citenamefont
  {Sangtawesin}, \citenamefont {McLellan}, \citenamefont {Myers}, \citenamefont
  {Jayich}, \citenamefont {Awschalom},\ and\ \citenamefont
  {Petta}}]{sangtawesin_hyperfine-enhanced_2016}%
  \BibitemOpen
  \bibfield  {author} {\bibinfo {author} {\bibfnamefont {S.}~\bibnamefont
  {Sangtawesin}}, \bibinfo {author} {\bibfnamefont {C.~A.}\ \bibnamefont
  {McLellan}}, \bibinfo {author} {\bibfnamefont {B.~A.}\ \bibnamefont {Myers}},
  \bibinfo {author} {\bibfnamefont {A.~C.~B.}\ \bibnamefont {Jayich}}, \bibinfo
  {author} {\bibfnamefont {D.~D.}\ \bibnamefont {Awschalom}},\ and\ \bibinfo
  {author} {\bibfnamefont {J.~R.}\ \bibnamefont {Petta}},\ }\bibfield  {title}
  {{\selectlanguage {en}\bibinfo {title} {Hyperfine-enhanced gyromagnetic ratio
  of a nuclear spin in diamond}},\ }\href
  {https://doi.org/10.1088/1367-2630/18/8/083016} {\bibfield  {journal}
  {\bibinfo  {journal} {New J. Phys.}\ }\textbf {\bibinfo {volume} {18}},\
  \bibinfo {pages} {083016} (\bibinfo {year} {2016})}\BibitemShut {NoStop}%
\bibitem [{\citenamefont {Degen}\ \emph {et~al.}(2021)\citenamefont {Degen},
  \citenamefont {Loenen}, \citenamefont {Bartling}, \citenamefont {Bradley},
  \citenamefont {Meinsma}, \citenamefont {Markham}, \citenamefont {Twitchen},\
  and\ \citenamefont {Taminiau}}]{degen_entanglement_2021}%
  \BibitemOpen
  \bibfield  {author} {\bibinfo {author} {\bibfnamefont {M.~J.}\ \bibnamefont
  {Degen}}, \bibinfo {author} {\bibfnamefont {S.~J.~H.}\ \bibnamefont
  {Loenen}}, \bibinfo {author} {\bibfnamefont {H.~P.}\ \bibnamefont
  {Bartling}}, \bibinfo {author} {\bibfnamefont {C.~E.}\ \bibnamefont
  {Bradley}}, \bibinfo {author} {\bibfnamefont {A.~L.}\ \bibnamefont
  {Meinsma}}, \bibinfo {author} {\bibfnamefont {M.}~\bibnamefont {Markham}},
  \bibinfo {author} {\bibfnamefont {D.~J.}\ \bibnamefont {Twitchen}},\ and\
  \bibinfo {author} {\bibfnamefont {T.~H.}\ \bibnamefont {Taminiau}},\
  }\bibfield  {title} {{\selectlanguage {en}\bibinfo {title} {Entanglement of
  dark electron-nuclear spin defects in diamond}},\ }\href
  {https://doi.org/10.1038/s41467-021-23454-9} {\bibfield  {journal} {\bibinfo
  {journal} {Nat Commun}\ }\textbf {\bibinfo {volume} {12}},\ \bibinfo {pages}
  {3470} (\bibinfo {year} {2021})}\BibitemShut {NoStop}%
\bibitem [{\citenamefont {Chen}\ \emph {et~al.}(2015)\citenamefont {Chen},
  \citenamefont {Hirose},\ and\ \citenamefont
  {Cappellaro}}]{chen_measurement_2015}%
  \BibitemOpen
  \bibfield  {author} {\bibinfo {author} {\bibfnamefont {M.}~\bibnamefont
  {Chen}}, \bibinfo {author} {\bibfnamefont {M.}~\bibnamefont {Hirose}},\ and\
  \bibinfo {author} {\bibfnamefont {P.}~\bibnamefont {Cappellaro}},\ }\bibfield
   {title} {\bibinfo {title} {Measurement of transverse hyperfine interaction
  by forbidden transitions},\ }\href
  {https://doi.org/10.1103/PhysRevB.92.020101} {\bibfield  {journal} {\bibinfo
  {journal} {Phys. Rev. B}\ }\textbf {\bibinfo {volume} {92}},\ \bibinfo
  {pages} {020101} (\bibinfo {year} {2015})}\BibitemShut {NoStop}%
\bibitem [{\citenamefont {Sangtawesin}(2016)}]{sangtawesin_quantum_2016}%
  \BibitemOpen
  \bibfield  {author} {\bibinfo {author} {\bibfnamefont {S.}~\bibnamefont
  {Sangtawesin}},\ }\emph {\bibinfo {title} {Quantum {Control} of {Nuclear}
  {Spins} {Coupled} to {Nitrogen}-{Vacancy} {Centers} in {Diamond}}},\ \href
  {https://ui.adsabs.harvard.edu/abs/2016PhDT........97S} {Ph.D. thesis}
  (\bibinfo {year} {2016})\BibitemShut {NoStop}%
\bibitem [{\citenamefont {Goldman}\ \emph {et~al.}(2020)\citenamefont
  {Goldman}, \citenamefont {Patti}, \citenamefont {Levonian}, \citenamefont
  {Yelin},\ and\ \citenamefont {Lukin}}]{goldman_optical_2020}%
  \BibitemOpen
  \bibfield  {author} {\bibinfo {author} {\bibfnamefont {M.}~\bibnamefont
  {Goldman}}, \bibinfo {author} {\bibfnamefont {T.}~\bibnamefont {Patti}},
  \bibinfo {author} {\bibfnamefont {D.}~\bibnamefont {Levonian}}, \bibinfo
  {author} {\bibfnamefont {S.}~\bibnamefont {Yelin}},\ and\ \bibinfo {author}
  {\bibfnamefont {M.}~\bibnamefont {Lukin}},\ }\bibfield  {title} {\bibinfo
  {title} {Optical {Control} of a {Single} {Nuclear} {Spin} in the {Solid}
  {State}},\ }\href {https://doi.org/10.1103/PhysRevLett.124.153203} {\bibfield
   {journal} {\bibinfo  {journal} {Phys. Rev. Lett.}\ }\textbf {\bibinfo
  {volume} {124}},\ \bibinfo {pages} {153203} (\bibinfo {year}
  {2020})}\BibitemShut {NoStop}%
\bibitem [{\citenamefont {Childress}\ \emph {et~al.}(2006)\citenamefont
  {Childress}, \citenamefont {Gurudev~Dutt}, \citenamefont {Taylor},
  \citenamefont {Zibrov}, \citenamefont {Jelezko}, \citenamefont {Wrachtrup},
  \citenamefont {Hemmer},\ and\ \citenamefont
  {Lukin}}]{childress_coherent_2006}%
  \BibitemOpen
  \bibfield  {author} {\bibinfo {author} {\bibfnamefont {L.}~\bibnamefont
  {Childress}}, \bibinfo {author} {\bibfnamefont {M.~V.}\ \bibnamefont
  {Gurudev~Dutt}}, \bibinfo {author} {\bibfnamefont {J.~M.}\ \bibnamefont
  {Taylor}}, \bibinfo {author} {\bibfnamefont {A.~S.}\ \bibnamefont {Zibrov}},
  \bibinfo {author} {\bibfnamefont {F.}~\bibnamefont {Jelezko}}, \bibinfo
  {author} {\bibfnamefont {J.}~\bibnamefont {Wrachtrup}}, \bibinfo {author}
  {\bibfnamefont {P.~R.}\ \bibnamefont {Hemmer}},\ and\ \bibinfo {author}
  {\bibfnamefont {M.~D.}\ \bibnamefont {Lukin}},\ }\bibfield  {title} {\bibinfo
  {title} {Coherent {Dynamics} of {Coupled} {Electron} and {Nuclear} {Spin}
  {Qubits} in {Diamond}},\ }\href {https://doi.org/10.1126/science.1131871}
  {\bibfield  {journal} {\bibinfo  {journal} {Science}\ }\textbf {\bibinfo
  {volume} {314}},\ \bibinfo {pages} {281} (\bibinfo {year}
  {2006})}\BibitemShut {NoStop}%
\bibitem [{\citenamefont {Jarmola}\ \emph {et~al.}(2020)\citenamefont
  {Jarmola}, \citenamefont {Fescenko}, \citenamefont {Acosta}, \citenamefont
  {Doherty}, \citenamefont {Fatemi}, \citenamefont {Ivanov}, \citenamefont
  {Budker},\ and\ \citenamefont {Malinovsky}}]{jarmola_robust_2020}%
  \BibitemOpen
  \bibfield  {author} {\bibinfo {author} {\bibfnamefont {A.}~\bibnamefont
  {Jarmola}}, \bibinfo {author} {\bibfnamefont {I.}~\bibnamefont {Fescenko}},
  \bibinfo {author} {\bibfnamefont {V.~M.}\ \bibnamefont {Acosta}}, \bibinfo
  {author} {\bibfnamefont {M.~W.}\ \bibnamefont {Doherty}}, \bibinfo {author}
  {\bibfnamefont {F.~K.}\ \bibnamefont {Fatemi}}, \bibinfo {author}
  {\bibfnamefont {T.}~\bibnamefont {Ivanov}}, \bibinfo {author} {\bibfnamefont
  {D.}~\bibnamefont {Budker}},\ and\ \bibinfo {author} {\bibfnamefont {V.~S.}\
  \bibnamefont {Malinovsky}},\ }\bibfield  {title} {\bibinfo {title} {Robust
  optical readout and characterization of nuclear spin transitions in
  nitrogen-vacancy ensembles in diamond},\ }\href
  {https://doi.org/10.1103/PhysRevResearch.2.023094} {\bibfield  {journal}
  {\bibinfo  {journal} {Phys. Rev. Research}\ }\textbf {\bibinfo {volume}
  {2}},\ \bibinfo {pages} {023094} (\bibinfo {year} {2020})}\BibitemShut
  {NoStop}%
\bibitem [{\citenamefont {Wood}\ \emph
  {et~al.}(2021{\natexlab{a}})\citenamefont {Wood}, \citenamefont {Goldblatt},
  \citenamefont {Scholten},\ and\ \citenamefont {Martin}}]{wood_quantum_2021}%
  \BibitemOpen
  \bibfield  {author} {\bibinfo {author} {\bibfnamefont {A.~A.}\ \bibnamefont
  {Wood}}, \bibinfo {author} {\bibfnamefont {R.~M.}\ \bibnamefont {Goldblatt}},
  \bibinfo {author} {\bibfnamefont {R.~E.}\ \bibnamefont {Scholten}},\ and\
  \bibinfo {author} {\bibfnamefont {A.~M.}\ \bibnamefont {Martin}},\ }\bibfield
   {title} {\bibinfo {title} {Quantum control of nuclear spin qubits in a
  rapidly rotating diamond},\ }\href {http://arxiv.org/abs/2107.12577}
  {\bibfield  {journal} {\bibinfo  {journal} {arXiv:2107.12577 [quant-ph]}\ }
  (\bibinfo {year} {2021}{\natexlab{a}})}\BibitemShut {NoStop}%
\bibitem [{\citenamefont {Laraoui}\ \emph {et~al.}(2012)\citenamefont
  {Laraoui}, \citenamefont {Hodges},\ and\ \citenamefont
  {Meriles}}]{laraoui_nitrogen-vacancy-assisted_2012}%
  \BibitemOpen
  \bibfield  {author} {\bibinfo {author} {\bibfnamefont {A.}~\bibnamefont
  {Laraoui}}, \bibinfo {author} {\bibfnamefont {J.~S.}\ \bibnamefont
  {Hodges}},\ and\ \bibinfo {author} {\bibfnamefont {C.~A.}\ \bibnamefont
  {Meriles}},\ }\bibfield  {title} {\bibinfo {title}
  {Nitrogen-{Vacancy}-{Assisted} {Magnetometry} of {Paramagnetic} {Centers} in
  an {Individual} {Diamond} {Nanocrystal}},\ }\href
  {https://doi.org/10.1021/nl300964g} {\bibfield  {journal} {\bibinfo
  {journal} {Nano Lett.}\ }\textbf {\bibinfo {volume} {12}},\ \bibinfo {pages}
  {3477} (\bibinfo {year} {2012})}\BibitemShut {NoStop}%
\bibitem [{\citenamefont {Belthangady}\ \emph {et~al.}(2013)\citenamefont
  {Belthangady}, \citenamefont {Bar-Gill}, \citenamefont {Pham}, \citenamefont
  {Arai}, \citenamefont {Le~Sage}, \citenamefont {Cappellaro},\ and\
  \citenamefont {Walsworth}}]{belthangady_dressed-state_2013}%
  \BibitemOpen
  \bibfield  {author} {\bibinfo {author} {\bibfnamefont {C.}~\bibnamefont
  {Belthangady}}, \bibinfo {author} {\bibfnamefont {N.}~\bibnamefont
  {Bar-Gill}}, \bibinfo {author} {\bibfnamefont {L.~M.}\ \bibnamefont {Pham}},
  \bibinfo {author} {\bibfnamefont {K.}~\bibnamefont {Arai}}, \bibinfo {author}
  {\bibfnamefont {D.}~\bibnamefont {Le~Sage}}, \bibinfo {author} {\bibfnamefont
  {P.}~\bibnamefont {Cappellaro}},\ and\ \bibinfo {author} {\bibfnamefont
  {R.~L.}\ \bibnamefont {Walsworth}},\ }\bibfield  {title} {\bibinfo {title}
  {Dressed-{State} {Resonant} {Coupling} between {Bright} and {Dark} {Spins} in
  {Diamond}},\ }\href {https://doi.org/10.1103/PhysRevLett.110.157601}
  {\bibfield  {journal} {\bibinfo  {journal} {Phys. Rev. Lett.}\ }\textbf
  {\bibinfo {volume} {110}},\ \bibinfo {pages} {157601} (\bibinfo {year}
  {2013})}\BibitemShut {NoStop}%
\bibitem [{\citenamefont {Knowles}\ \emph {et~al.}(2016)\citenamefont
  {Knowles}, \citenamefont {Kara},\ and\ \citenamefont
  {Atatüre}}]{knowles_demonstration_2016}%
  \BibitemOpen
  \bibfield  {author} {\bibinfo {author} {\bibfnamefont {H.~S.}\ \bibnamefont
  {Knowles}}, \bibinfo {author} {\bibfnamefont {D.~M.}\ \bibnamefont {Kara}},\
  and\ \bibinfo {author} {\bibfnamefont {M.}~\bibnamefont {Atatüre}},\
  }\bibfield  {title} {\bibinfo {title} {Demonstration of a {Coherent}
  {Electronic} {Spin} {Cluster} in {Diamond}},\ }\href
  {https://doi.org/10.1103/PhysRevLett.117.100802} {\bibfield  {journal}
  {\bibinfo  {journal} {Phys. Rev. Lett.}\ }\textbf {\bibinfo {volume} {117}},\
  \bibinfo {pages} {100802} (\bibinfo {year} {2016})}\BibitemShut {NoStop}%
\bibitem [{\citenamefont {de~Lange}\ \emph {et~al.}(2012)\citenamefont
  {de~Lange}, \citenamefont {van~der Sar}, \citenamefont {Blok}, \citenamefont
  {Wang}, \citenamefont {Dobrovitski},\ and\ \citenamefont
  {Hanson}}]{de_lange_controlling_2012}%
  \BibitemOpen
  \bibfield  {author} {\bibinfo {author} {\bibfnamefont {G.}~\bibnamefont
  {de~Lange}}, \bibinfo {author} {\bibfnamefont {T.}~\bibnamefont {van~der
  Sar}}, \bibinfo {author} {\bibfnamefont {M.}~\bibnamefont {Blok}}, \bibinfo
  {author} {\bibfnamefont {Z.-H.}\ \bibnamefont {Wang}}, \bibinfo {author}
  {\bibfnamefont {V.}~\bibnamefont {Dobrovitski}},\ and\ \bibinfo {author}
  {\bibfnamefont {R.}~\bibnamefont {Hanson}},\ }\bibfield  {title}
  {{\selectlanguage {en}\bibinfo {title} {Controlling the quantum dynamics of a
  mesoscopic spin bath in diamond}},\ }\href
  {https://doi.org/10.1038/srep00382} {\bibfield  {journal} {\bibinfo
  {journal} {Sci Rep}\ }\textbf {\bibinfo {volume} {2}},\ \bibinfo {pages}
  {382} (\bibinfo {year} {2012})}\BibitemShut {NoStop}%
\bibitem [{\citenamefont {Knowles}\ \emph {et~al.}(2014)\citenamefont
  {Knowles}, \citenamefont {Kara},\ and\ \citenamefont
  {Atatüre}}]{knowles_observing_2014}%
  \BibitemOpen
  \bibfield  {author} {\bibinfo {author} {\bibfnamefont {H.~S.}\ \bibnamefont
  {Knowles}}, \bibinfo {author} {\bibfnamefont {D.~M.}\ \bibnamefont {Kara}},\
  and\ \bibinfo {author} {\bibfnamefont {M.}~\bibnamefont {Atatüre}},\
  }\bibfield  {title} {{\selectlanguage {en}\bibinfo {title} {Observing bulk
  diamond spin coherence in high-purity nanodiamonds}},\ }\href
  {https://doi.org/10.1038/nmat3805} {\bibfield  {journal} {\bibinfo  {journal}
  {Nature Mater}\ }\textbf {\bibinfo {volume} {13}},\ \bibinfo {pages} {21}
  (\bibinfo {year} {2014})}\BibitemShut {NoStop}%
\bibitem [{\citenamefont {Bauch}\ \emph {et~al.}(2020)\citenamefont {Bauch},
  \citenamefont {Singh}, \citenamefont {Lee}, \citenamefont {Hart},
  \citenamefont {Schloss}, \citenamefont {Turner}, \citenamefont {Barry},
  \citenamefont {Pham}, \citenamefont {Bar-Gill}, \citenamefont {Yelin},\ and\
  \citenamefont {Walsworth}}]{bauch_decoherence_2020}%
  \BibitemOpen
  \bibfield  {author} {\bibinfo {author} {\bibfnamefont {E.}~\bibnamefont
  {Bauch}}, \bibinfo {author} {\bibfnamefont {S.}~\bibnamefont {Singh}},
  \bibinfo {author} {\bibfnamefont {J.}~\bibnamefont {Lee}}, \bibinfo {author}
  {\bibfnamefont {C.~A.}\ \bibnamefont {Hart}}, \bibinfo {author}
  {\bibfnamefont {J.~M.}\ \bibnamefont {Schloss}}, \bibinfo {author}
  {\bibfnamefont {M.~J.}\ \bibnamefont {Turner}}, \bibinfo {author}
  {\bibfnamefont {J.~F.}\ \bibnamefont {Barry}}, \bibinfo {author}
  {\bibfnamefont {L.~M.}\ \bibnamefont {Pham}}, \bibinfo {author}
  {\bibfnamefont {N.}~\bibnamefont {Bar-Gill}}, \bibinfo {author}
  {\bibfnamefont {S.~F.}\ \bibnamefont {Yelin}},\ and\ \bibinfo {author}
  {\bibfnamefont {R.~L.}\ \bibnamefont {Walsworth}},\ }\bibfield  {title}
  {\bibinfo {title} {Decoherence of ensembles of nitrogen-vacancy centers in
  diamond},\ }\href {https://doi.org/10.1103/PhysRevB.102.134210} {\bibfield
  {journal} {\bibinfo  {journal} {Phys. Rev. B}\ }\textbf {\bibinfo {volume}
  {102}},\ \bibinfo {pages} {134210} (\bibinfo {year} {2020})}\BibitemShut
  {NoStop}%
\bibitem [{\citenamefont {de~Lange}\ \emph {et~al.}(2010)\citenamefont
  {de~Lange}, \citenamefont {Wang}, \citenamefont {Ristè}, \citenamefont
  {Dobrovitski},\ and\ \citenamefont {Hanson}}]{de_lange_universal_2010}%
  \BibitemOpen
  \bibfield  {author} {\bibinfo {author} {\bibfnamefont {G.}~\bibnamefont
  {de~Lange}}, \bibinfo {author} {\bibfnamefont {Z.~H.}\ \bibnamefont {Wang}},
  \bibinfo {author} {\bibfnamefont {D.}~\bibnamefont {Ristè}}, \bibinfo
  {author} {\bibfnamefont {V.~V.}\ \bibnamefont {Dobrovitski}},\ and\ \bibinfo
  {author} {\bibfnamefont {R.}~\bibnamefont {Hanson}},\ }\bibfield  {title}
  {\bibinfo {title} {Universal {Dynamical} {Decoupling} of a {Single}
  {Solid}-{State} {Spin} from a {Spin} {Bath}},\ }\href
  {https://doi.org/10.1126/science.1192739} {\bibfield  {journal} {\bibinfo
  {journal} {Science}\ }\textbf {\bibinfo {volume} {330}},\ \bibinfo {pages}
  {60} (\bibinfo {year} {2010})}\BibitemShut {NoStop}%
\bibitem [{\citenamefont {Hansom}\ \emph {et~al.}(2014)\citenamefont {Hansom},
  \citenamefont {Schulte}, \citenamefont {Le~Gall}, \citenamefont {Matthiesen},
  \citenamefont {Clarke}, \citenamefont {Hugues}, \citenamefont {Taylor},\ and\
  \citenamefont {Atatüre}}]{hansom_environment-assisted_2014}%
  \BibitemOpen
  \bibfield  {author} {\bibinfo {author} {\bibfnamefont {J.}~\bibnamefont
  {Hansom}}, \bibinfo {author} {\bibfnamefont {C.~H.~H.}\ \bibnamefont
  {Schulte}}, \bibinfo {author} {\bibfnamefont {C.}~\bibnamefont {Le~Gall}},
  \bibinfo {author} {\bibfnamefont {C.}~\bibnamefont {Matthiesen}}, \bibinfo
  {author} {\bibfnamefont {E.}~\bibnamefont {Clarke}}, \bibinfo {author}
  {\bibfnamefont {M.}~\bibnamefont {Hugues}}, \bibinfo {author} {\bibfnamefont
  {J.~M.}\ \bibnamefont {Taylor}},\ and\ \bibinfo {author} {\bibfnamefont
  {M.}~\bibnamefont {Atatüre}},\ }\bibfield  {title} {{\selectlanguage
  {en}\bibinfo {title} {Environment-assisted quantum control of a solid-state
  spin via coherent dark states}},\ }\href {https://doi.org/10.1038/nphys3077}
  {\bibfield  {journal} {\bibinfo  {journal} {Nature Phys}\ }\textbf {\bibinfo
  {volume} {10}},\ \bibinfo {pages} {725} (\bibinfo {year} {2014})}\BibitemShut
  {NoStop}%
\bibitem [{\citenamefont {Bauch}\ \emph {et~al.}(2018)\citenamefont {Bauch},
  \citenamefont {Hart}, \citenamefont {Schloss}, \citenamefont {Turner},
  \citenamefont {Barry}, \citenamefont {Kehayias}, \citenamefont {Singh},\ and\
  \citenamefont {Walsworth}}]{bauch_ultralong_2018}%
  \BibitemOpen
  \bibfield  {author} {\bibinfo {author} {\bibfnamefont {E.}~\bibnamefont
  {Bauch}}, \bibinfo {author} {\bibfnamefont {C.~A.}\ \bibnamefont {Hart}},
  \bibinfo {author} {\bibfnamefont {J.~M.}\ \bibnamefont {Schloss}}, \bibinfo
  {author} {\bibfnamefont {M.~J.}\ \bibnamefont {Turner}}, \bibinfo {author}
  {\bibfnamefont {J.~F.}\ \bibnamefont {Barry}}, \bibinfo {author}
  {\bibfnamefont {P.}~\bibnamefont {Kehayias}}, \bibinfo {author}
  {\bibfnamefont {S.}~\bibnamefont {Singh}},\ and\ \bibinfo {author}
  {\bibfnamefont {R.~L.}\ \bibnamefont {Walsworth}},\ }\bibfield  {title}
  {\bibinfo {title} {Ultralong {Dephasing} {Times} in {Solid}-{State} {Spin}
  {Ensembles} via {Quantum} {Control}},\ }\bibfield  {journal} {\bibinfo
  {journal} {Phys. Rev. X}\ }\textbf {\bibinfo {volume} {8}},\ \href
  {https://doi.org/10.1103/PhysRevX.8.031025} {10.1103/PhysRevX.8.031025}
  (\bibinfo {year} {2018})\BibitemShut {NoStop}%
\bibitem [{\citenamefont {Wood}\ \emph
  {et~al.}(2021{\natexlab{b}})\citenamefont {Wood}, \citenamefont {Goldblatt},
  \citenamefont {Anderson}, \citenamefont {Hollenberg}, \citenamefont
  {Scholten},\ and\ \citenamefont {Martin}}]{wood_anisotropic_2021}%
  \BibitemOpen
  \bibfield  {author} {\bibinfo {author} {\bibfnamefont {A.~A.}\ \bibnamefont
  {Wood}}, \bibinfo {author} {\bibfnamefont {R.~M.}\ \bibnamefont {Goldblatt}},
  \bibinfo {author} {\bibfnamefont {R.~P.}\ \bibnamefont {Anderson}}, \bibinfo
  {author} {\bibfnamefont {L.~C.~L.}\ \bibnamefont {Hollenberg}}, \bibinfo
  {author} {\bibfnamefont {R.~E.}\ \bibnamefont {Scholten}},\ and\ \bibinfo
  {author} {\bibfnamefont {A.~M.}\ \bibnamefont {Martin}},\ }\bibfield  {title}
  {\bibinfo {title} {Anisotropic electron-nuclear interactions in a rotating
  quantum spin bath},\ }\href {https://doi.org/10.1103/PhysRevB.104.085419}
  {\bibfield  {journal} {\bibinfo  {journal} {Phys. Rev. B}\ }\textbf {\bibinfo
  {volume} {104}},\ \bibinfo {pages} {085419} (\bibinfo {year}
  {2021}{\natexlab{b}})}\BibitemShut {NoStop}%
\bibitem [{Note1()}]{Note1}%
  \BibitemOpen
  \bibinfo {note} {See Supplementary Information.}\BibitemShut {Stop}%
\bibitem [{Note2()}]{Note2}%
  \BibitemOpen
  \bibinfo {note} {See Supplementary information}\BibitemShut {NoStop}%
\bibitem [{\citenamefont {Davies}(1979)}]{davies_dynamic_1979}%
  \BibitemOpen
  \bibfield  {author} {\bibinfo {author} {\bibfnamefont {G.}~\bibnamefont
  {Davies}},\ }\bibfield  {title} {{\selectlanguage {en}\bibinfo {title}
  {Dynamic {Jahn}-{Teller} distortions at trigonal optical centres in
  diamond}},\ }\href {https://doi.org/10.1088/0022-3719/12/13/019} {\bibfield
  {journal} {\bibinfo  {journal} {J. Phys. C: Solid State Phys.}\ }\textbf
  {\bibinfo {volume} {12}},\ \bibinfo {pages} {2551} (\bibinfo {year}
  {1979})}\BibitemShut {NoStop}%
\bibitem [{\citenamefont {Davies}(1981)}]{davies_jahn-teller_1981}%
  \BibitemOpen
  \bibfield  {author} {\bibinfo {author} {\bibfnamefont {G.}~\bibnamefont
  {Davies}},\ }\bibfield  {title} {{\selectlanguage {en}\bibinfo {title} {The
  {Jahn}-{Teller} effect and vibronic coupling at deep levels in diamond}},\
  }\href {https://doi.org/10.1088/0034-4885/44/7/003} {\bibfield  {journal}
  {\bibinfo  {journal} {Rep. Prog. Phys.}\ }\textbf {\bibinfo {volume} {44}},\
  \bibinfo {pages} {787} (\bibinfo {year} {1981})}\BibitemShut {NoStop}%
\bibitem [{\citenamefont {Ammerlaan}\ and\ \citenamefont
  {Burgemeister}(1981)}]{ammerlaan_reorientation_1981}%
  \BibitemOpen
  \bibfield  {author} {\bibinfo {author} {\bibfnamefont {C.~A.~J.}\
  \bibnamefont {Ammerlaan}}\ and\ \bibinfo {author} {\bibfnamefont {E.~A.}\
  \bibnamefont {Burgemeister}},\ }\bibfield  {title} {\bibinfo {title}
  {Reorientation of {Nitrogen} in {Type}-{$\mathrm{I}b$} {Diamond} by {Thermal}
  {Excitation} and {Tunneling}},\ }\href
  {https://doi.org/10.1103/PhysRevLett.47.954} {\bibfield  {journal} {\bibinfo
  {journal} {Phys. Rev. Lett.}\ }\textbf {\bibinfo {volume} {47}},\ \bibinfo
  {pages} {954} (\bibinfo {year} {1981})}\BibitemShut {NoStop}%
\bibitem [{\citenamefont {Larsen}\ and\ \citenamefont
  {Singel}(1993)}]{larsen_double_1993}%
  \BibitemOpen
  \bibfield  {author} {\bibinfo {author} {\bibfnamefont {R.~G.}\ \bibnamefont
  {Larsen}}\ and\ \bibinfo {author} {\bibfnamefont {D.~J.}\ \bibnamefont
  {Singel}},\ }\bibfield  {title} {\bibinfo {title} {Double electron–electron
  resonance spin–echo modulation: {Spectroscopic} measurement of electron
  spin pair separations in orientationally disordered solids},\ }\href
  {https://doi.org/10.1063/1.464916} {\bibfield  {journal} {\bibinfo  {journal}
  {J. Chem. Phys.}\ }\textbf {\bibinfo {volume} {98}},\ \bibinfo {pages} {5134}
  (\bibinfo {year} {1993})}\BibitemShut {NoStop}%
\bibitem [{\citenamefont {Pagliero}\ \emph {et~al.}(2018)\citenamefont
  {Pagliero}, \citenamefont {Rao}, \citenamefont {Zangara}, \citenamefont
  {Dhomkar}, \citenamefont {Wong}, \citenamefont {Abril}, \citenamefont
  {Aslam}, \citenamefont {Parker}, \citenamefont {King}, \citenamefont
  {Avalos}, \citenamefont {Ajoy}, \citenamefont {Wrachtrup}, \citenamefont
  {Pines},\ and\ \citenamefont {Meriles}}]{pagliero_multispin-assisted_2018}%
  \BibitemOpen
  \bibfield  {author} {\bibinfo {author} {\bibfnamefont {D.}~\bibnamefont
  {Pagliero}}, \bibinfo {author} {\bibfnamefont {K.~R.~K.}\ \bibnamefont
  {Rao}}, \bibinfo {author} {\bibfnamefont {P.~R.}\ \bibnamefont {Zangara}},
  \bibinfo {author} {\bibfnamefont {S.}~\bibnamefont {Dhomkar}}, \bibinfo
  {author} {\bibfnamefont {H.~H.}\ \bibnamefont {Wong}}, \bibinfo {author}
  {\bibfnamefont {A.}~\bibnamefont {Abril}}, \bibinfo {author} {\bibfnamefont
  {N.}~\bibnamefont {Aslam}}, \bibinfo {author} {\bibfnamefont
  {A.}~\bibnamefont {Parker}}, \bibinfo {author} {\bibfnamefont
  {J.}~\bibnamefont {King}}, \bibinfo {author} {\bibfnamefont {C.~E.}\
  \bibnamefont {Avalos}}, \bibinfo {author} {\bibfnamefont {A.}~\bibnamefont
  {Ajoy}}, \bibinfo {author} {\bibfnamefont {J.}~\bibnamefont {Wrachtrup}},
  \bibinfo {author} {\bibfnamefont {A.}~\bibnamefont {Pines}},\ and\ \bibinfo
  {author} {\bibfnamefont {C.~A.}\ \bibnamefont {Meriles}},\ }\bibfield
  {title} {\bibinfo {title} {Multispin-assisted optical pumping of bulk
  {$^{13}\mathrm{C}$} nuclear spin polarization in diamond},\ }\href
  {https://doi.org/10.1103/PhysRevB.97.024422} {\bibfield  {journal} {\bibinfo
  {journal} {Phys. Rev. B}\ }\textbf {\bibinfo {volume} {97}},\ \bibinfo
  {pages} {024422} (\bibinfo {year} {2018})}\BibitemShut {NoStop}%
\bibitem [{Note3()}]{Note3}%
  \BibitemOpen
  \bibinfo {note} {See Supplemental information}\BibitemShut {NoStop}%
\bibitem [{\citenamefont {Childress}\ and\ \citenamefont
  {Hanson}(2013)}]{childress_diamond_2013}%
  \BibitemOpen
  \bibfield  {author} {\bibinfo {author} {\bibfnamefont {L.}~\bibnamefont
  {Childress}}\ and\ \bibinfo {author} {\bibfnamefont {R.}~\bibnamefont
  {Hanson}},\ }\bibfield  {title} {{\selectlanguage {en}\bibinfo {title}
  {Diamond {NV} centers for quantum computing and quantum networks}},\ }\href
  {https://doi.org/10.1557/mrs.2013.20} {\bibfield  {journal} {\bibinfo
  {journal} {MRS Bulletin}\ }\textbf {\bibinfo {volume} {38}},\ \bibinfo
  {pages} {134} (\bibinfo {year} {2013})}\BibitemShut {NoStop}%
\end{thebibliography}

\end{document}